\documentclass[preprint,flushrt]{aastex}
\usepackage{natbib}
\usepackage{lineno}
\newcommand\pflux{\mbox{${\rm \, ph \,\, cm^{-2} \, s^{-1}}$}}
\linenumbers

\shorttitle{The radio/gamma-ray connection in AGNs in the {\it Fermi} LAT era}
\shortauthors{Ackermann et al.}

\title{The radio/gamma-ray connection in Active Galactic Nuclei in the era of
  the {\it Fermi} Large Area Telescope}

\author{
M.~Ackermann\altaffilmark{2}, 
M.~Ajello\altaffilmark{2}, 
A.~Allafort\altaffilmark{2}, 
E.~Angelakis\altaffilmark{3}, 
M.~Axelsson\altaffilmark{4,5,6}, 
L.~Baldini\altaffilmark{7}, 
J.~Ballet\altaffilmark{8}, 
G.~Barbiellini\altaffilmark{9,10}, 
D.~Bastieri\altaffilmark{11,12}, 
R.~Bellazzini\altaffilmark{7}
% ...
}
% ...
\altaffiltext{2}{W. W. Hansen Experimental Physics Laboratory, Kavli Institute for Particle Astrophysics and Cosmology, Department of Physics and SLAC National Accelerator Laboratory, Stanford University, Stanford, CA 94305, USA}
\altaffiltext{3}{Max-Planck-Institut f\"ur Radioastronomie, Auf dem H\"ugel 69, 53121 Bonn, Germany}
\altaffiltext{4}{Department of Astronomy, Stockholm University, SE-106 91 Stockholm, Sweden}
\altaffiltext{5}{Lund Observatory, SE-221 00 Lund, Sweden}
\altaffiltext{6}{The Oskar Klein Centre for Cosmoparticle Physics, AlbaNova, SE-106 91 Stockholm, Sweden}
\altaffiltext{7}{Istituto Nazionale di Fisica Nucleare, Sezione di Pisa, I-56127 Pisa, Italy}
\altaffiltext{8}{Laboratoire AIM, CEA-IRFU/CNRS/Universit\'e Paris Diderot, Service d'Astrophysique, CEA Saclay, 91191 Gif sur Yvette, France}
\altaffiltext{9}{Istituto Nazionale di Fisica Nucleare, Sezione di Trieste, I-34127 Trieste, Italy}
\altaffiltext{10}{Dipartimento di Fisica, Universit\`a di Trieste, I-34127 Trieste, Italy}
\altaffiltext{11}{Istituto Nazionale di Fisica Nucleare, Sezione di Padova, I-35131 Padova, Italy}
\altaffiltext{12}{Dipartimento di Fisica ``G. Galilei", Universit\`a di Padova, I-35131 Padova, Italy}
% ...

\begin{abstract}

We present a detailed statistical analysis of the correlation between radio and
gamma-ray emission of the Active Galactic Nuclei (AGN) detected by {\it Fermi}
during its first year of operation, with the largest datasets ever used for
this purpose.  We use both archival interferometric 8.4 GHz data (from the VLA
and ATCA, for the full sample of 599 sources) and concurrent single-dish 15 GHz
measurements from the Owens Valley Radio Observatory (OVRO, for a sub sample of
199 objects). Our unprecedentedly large sample permits us to assess with high
accuracy the statistical significance of the correlation, using a
surrogate-data method designed to simultaneously account for common-distance
bias and the effect of a limited dynamical range in the observed quantities.
We find that the statistical significance of a positive correlation between the
cm radio and the broad band ($E>100\,$MeV) gamma-ray energy flux is very high
for the whole AGN sample, with a probability $<10^{-7}$ for the correlation
appearing by chance. Using the OVRO data, we find that concurrent data improve
the significance of the correlation from $1.6 \times 10^{-6}$ to $9.0 \times
10^{-8}$.  Our large sample size allows us to study the dependence of
correlation strength and significance on specific source types and gamma-ray
energy band. We find that the correlation is very significant (chance
probability $<10^{-7}$) for both FSRQs and BL Lacs separately; a dependence of
the correlation strength on the considered gamma-ray energy band is also
present, but additional data will be necessary to constrain its significance.

\end{abstract}


\begin{document}

\section{Introduction}

After more than 1 year of scanning the gamma-ray sky by the Large Area
Telescope (LAT) onboard the {\it Fermi Gamma-ray Space Telescope (Fermi)}, the
most extreme class of Active Galactic Nuclei (AGN), blazars (used to refer
collectively to BL~Lac objects, hereafter BL Lacs, and flat spectrum radio
quasars, hereafter FSRQs), remain among the most numerous gamma-ray source
populations.  Indeed, the First {\it Fermi}-LAT catalog of gamma-ray sources
\citep[hereafter 1FGL,][]{Catalog} includes more than 1400 sources and about
half of them are believed to be AGNs \citep{1LAC} with most of them identified
via radio catalogs \citep[e.g., CRATES;][]{crates}.  More than 370
high-latitude ($|b|>10^\circ$) sources in the 1FGL remain unidentified.

Blazars have been observed to emit at all energies, from the radio band up to
very-high energy gamma-rays.  Many of the gamma-ray blazars detected so far
appear to emit the bulk of their total radiative output at gamma-ray energies.
Strong variability across the whole electromagnetic spectrum and on various
time scales is considered as one of the most intriguing properties of this
source type. In particular their high-energy emission can easily vary by more
than an order of magnitude from one observing epoch to the next
\citep[e.g.][]{Mukherjee1997,LBASvar}, and variability time scales at high
energies are mostly much shorter \citep[even down to just a few minutes in the
  TeV band, e.g.][]{Aharonian2007} than in the long wavelength bands.

The high inferred bolometric luminosities, rapid variability, and apparent
superluminal motions observed from a range of blazars provide compelling
evidence that the non-thermal emission of blazars originates from a region
which is propagating relativistically along a jet directed at a small angle
with respect to our line of sight.

Because most identified gamma-ray AGN are classified as radio-loud objects, a
luminosity correlation between those two wavebands appears possible.  If proved
true, constraints on the physics and location of the jet emission from such AGN
may be deduced.  Many attempts have been made in the past to investigate
correlations between radio (cm)- and gamma-ray luminosities of AGN
\citep[e.g.,][]{Stecker1993,Padovani1993,Salamon1994,Taylor2007}. However, the
relation has not been conclusively demonstrated when all relevant biases and
selection effects are taken into account \citep[see, e.g.][]{Muecke1997}.

For example, while luminosities represent the intrinsic source property, as
opposed to fluxes, the use of luminosities always introduces a redshift bias in
samples which cover a wide distance range since luminosities are strongly
correlated with redshift \citep{Elvis1978}. Such redshift dependence can be
removed by means of a partial correlation analysis \citep[see
  e.g.,][]{Dondi1995}.  On the other hand, intrinsic correlations between the
gamma-ray and radio luminosities may be smeared out, or even lost in the
corresponding flux diagrams whereas artificial flux-flux correlations can be
induced due to the effect of a common distance modulation of gamma-ray and
radio luminosities \citep[the ``common-distance'' bias, see
  e.g.][]{Methodology}.

Samples that are strongly sensitivity limited restrict the populated region in
the luminosity-luminosity diagram to a narrow band, thereby causing serious
biases.  Therefore, \citet{Feigelson1983} proposed to include all upper limits
to avoid artificial correlations and incorrect conclusions \citep{Schmitt1985}.
However, upper limits are usually not distributed randomly in the flux-flux or
luminosity-luminosity plane, but are localized in a particular area.  In this
case, a survival analysis may give misleading results
\citep{Isobe1989}. Furthermore, this analysis cannot account for biases caused
by misidentification of sources or by truncation effects. Finally, the use of
rank correlation tests (e.g., Kendall's $\tau$, Spearman rank correlation
coefficient $\rho$) complicates the inclusion of observational uncertainties.

Another problem is the data and source selection.  Blazars are inherently
variable sources in the radio as well as the gamma-ray band on a broad range of
time scales.  Simultaneous observations are therefore the only appropriate data
for a correlation analysis. However, due to the lack of such data, the mean
\citep[e.g.,][]{Padovani1993} or the brightest flux values
\citep[e.g.,][]{Dondi1995} have often been used instead. As a consequence the
dynamical range in the luminosity-luminosity plane is significantly reduced in
those cases, and can hence mimic a correlation \citep{Muecke1997}.

The question of a correlation between the radio and GeV band on the basis of
{\em Fermi} data has recently generated a lot of interest and has been the
subject of a series of investigations
\citep{Kovalev2009,Ghirlanda2010,Ghirlanda2011,Mahony2010}. However, these
studies have been generally limited to a small fraction of the {\it
  Fermi}-detected AGN and have used non-simultaneous or quasi-simultaneous
measurements.  Moreover, these works have primarily addressed the issue of the
{\it apparent strength} of the correlation, rather than that of its {\it
  intrinsic significance}, which requires a dedicated method of statistical
analysis. In this paper, we will use the term "apparent correlation strength"
for measures of the tightness of a correlation between radio and gamma-ray {\em
  fluxes} (such as various correlation coefficients) as seen in the raw data,
without applying any correction or significance assessment to address
common-distance bias and the limits on the measured fluxes (the issue of
"censored data"). In contrast, we will use the term "intrinsic correlation" for
the physical correlation between radio and gamma-ray (time-averaged)
luminosities, in the limit of an infinite survey, and "intrinsic correlation
significance" for the statistical significance of the claim that a specific
dataset exhibits a non-zero intrinsic correlation.

In this paper, we revisit this topic exploiting for the first time the {\em
  Fermi}-LAT data in full, in two ways. First, we make use of archival data for
about 600 sources, a dataset more than twice as large as that used in
\citet{Ghirlanda2010} and \citet{Mahony2010}. Second, we take advantage of the
large set of concurrent measurements provided by the OVRO monitoring program
\citep{OVRO}.  The pre-{\em Fermi}-launch OVRO sample included $\sim 200$
blazars that are included in the 1FGL catalog, and for which average 15 GHz
fluxes measured {\em concurrently} with the 1FGL gamma-ray fluxes can be
calculated. In addition, we exploit a new statistical method
\citep{Methodology} to assess the significance of the correlation coefficients.
 
The paper is structured as follows: in Sect.~\ref{s.data}, we present the
gamma-ray and radio data and the association procedure; the results are
presented in Sect.~\ref{s.results} and discussed in Sect.~\ref{s.mc} using a
dedicated statistical analysis based on the method of surrogate data. A more
general discussion is given in Sect.~\ref{s.discussion} and the main
conclusions are summarized in Sect.~\ref{s.conclusions}.

In the following, we use a $\Lambda$CDM cosmology with 
$h = 0.71$, $\Omega_m = 0.27$, and $\Omega_\Lambda=0.73$ \citep{komatsu}.
The radio spectral index is defined such that $S(\nu) \propto \nu^{-\alpha}$
and the gamma-ray  photon
index $\Gamma$ such that  $dN_{\rm photon}/dE \propto E^{-\Gamma}$.

\section{Observations and dataset \label{s.data}}

\subsection{Gamma-ray data}

The \mbox{gamma-ray} sources in the present paper are a subset of those in the
First {\it Fermi}-LAT catalog \citep[1FGL,][]{Catalog}. The 1FGL is a catalog
of high-energy gamma-ray sources detected by the LAT during the first 11 months
of the science phase of {\it Fermi}, i.e.\ between 2008\,August\,4 and
2009\,July\,4.  The procedures used in producing the 1FGL catalog are discussed
in detail in \cite{Catalog}; in total, the 1FGL contains 1451 sources detected
and characterized in the 100 MeV to 100 GeV range and belonging to a number of
populations of gamma-ray emitters.

In general, associations of \mbox{gamma-ray} sources with lower-energy
counterparts necessarily rely on a spatial coincidence between the two. A firm
counterpart identification requires the search for correlated variability,
which is a major effort in the case of AGNs; therefore, only 5 AGNs are listed
as firm identifications by \citet{Catalog}, although ongoing studies will
undoubtedly expand this set.  For the rest, associations in 1FGL use a method
for finding correspondence between LAT sources and AGNs based on the
calculation of association probabilities using a Bayesian approach implemented
in the {\it gtsrcid} tool included in the LAT {\it ScienceTools} package.  A
detailed description and a complete list of the source catalogs used by {\it
  gtsrcid} to draw candidate counterparts can be found in \citet{Catalog}.

The set of all high-latitude ($|b| > 10\arcdeg$) 1FGL sources with an AGN
association from {\it gtsrcid} constitutes the First LAT AGN Catalog
\citep[1LAC,][]{1LAC}. Some LAT sources are associated with multiple AGNs, and
consequently, the catalog includes 709 AGN associations for 671 distinct 1FGL
sources. Each source has an association probability $P$, evaluated by examining
the local density of counterparts from a number of source catalogs in the
vicinity of the LAT source. The main catalogs used are the Combined Radio
All-sky Targeted Eight GHz Survey \citep[CRATES;][]{crates}, the Candidate
Gamma-Ray Blazar Survey \citep[CGRaBS;][]{cgrabs}, and the Roma-BZCAT
\citep{bzcat}. Since a few gamma-ray sources have more than one possible
association, and not all associations are highly significant, \citet{1LAC} have
further defined an AGN ``clean'' sample consisting of those AGNs that (1) are
the sole AGN associated with the corresponding 1FGL \mbox{gamma-ray} source and
(2) have an association probability $P \ge 80$\%; a few sources, ``flagged'' in
the 1FGL catalog as exhibiting some problem, have also been discarded and do
not belong in the 1LAC clean sample.  This clean sample contains 599 AGNs. In
the following analysis, whenever we mention the 1LAC sample, we will always be
referring to the clean sample even if we do not state so explicitly.

For each source in the 1FGL (and hence in the 1LAC), \citet{Catalog} have first
obtained good estimates of the significance and the overall spectral slope
$\Gamma$. Then, in order to obtain good estimates of the energy flux, each of
the five energy bands (from 100 to 300 MeV, 300 MeV to 1 GeV, 1 to 3 GeV, 3 to
10 GeV, and 10 to 100 GeV) has been fit independently, fixing the spectral
index of each source to $\Gamma$ as derived from the fit over the full
interval; finally, the sum of the energy flux in the five bands provided a
reliable estimate of the overall flux.

In sources with a poorly measured flux (88/599), \citet{Catalog} replaced the
value from the likelihood analysis with a $2\sigma$ upper limit. However, since
these sources are significantly detected when the full band is considered, we
estimated their energy fluxes from the flux densities at the pivot energies
given by \citet{Catalog}, and using the tabulated photon indices and the
relative uncertainties on the corresponding quantities. All the obtained data
are consistent with the $2\sigma$ limits and so have been used for our
analysis.

We maintain the 1LAC classification of each AGN on the basis of its optical
spectrum either as an FSRQ or a BL~Lac using the same scheme as in CGRaBS
\citep{cgrabs}.  In particular, following \citet{stocke91}, \citet{up95}, and
\citet{marcha96}, an object is classified as a BL~Lac if the equivalent width
(EW) of the strongest optical emission line is $< 5$~\AA, the optical spectrum
shows a \ion{Ca}{2} H/K break ratio $C < 0.4$, and the wavelength coverage of
the spectrum satisfies $(\lambda_\mathrm{max} -
\lambda_\mathrm{min})/\lambda_\mathrm{max} > 1.7$ in order to ensure that at
least one strong emission line would have been detected if it were present.

In addition to the optical spectrum classification, the 1LAC blazars are also
classified based on the position of the synchrotron peak, following the scheme
proposed by \citet{SEDpaper}; we therefore consider also the three following
spectral types: low-synchrotron-peaked blazars (LSP,
$\nu_\mathrm{peak}^\mathrm{S} < 10^{14}$~Hz); intermediate-synchrotron-peaked
(ISP, $10^{14}$~Hz~$< \nu_\mathrm{peak}^\mathrm{S} < 10^{15}$~Hz); or
high-synchrotron-peaked (HSP, $\nu_\mathrm{peak}^\mathrm{S} > 10^{15}$~Hz).
Althought the two classification schemes do have some degeneracy (e.g., HSP
sources are largely BL Lacs, while most FSRQs are LSP sources), it is relevant
to discuss them both, as the spectral classification is linked to the physical
process (synchrotron radiation) responsible for the low frequency emission.

\begin{figure}
\plotone{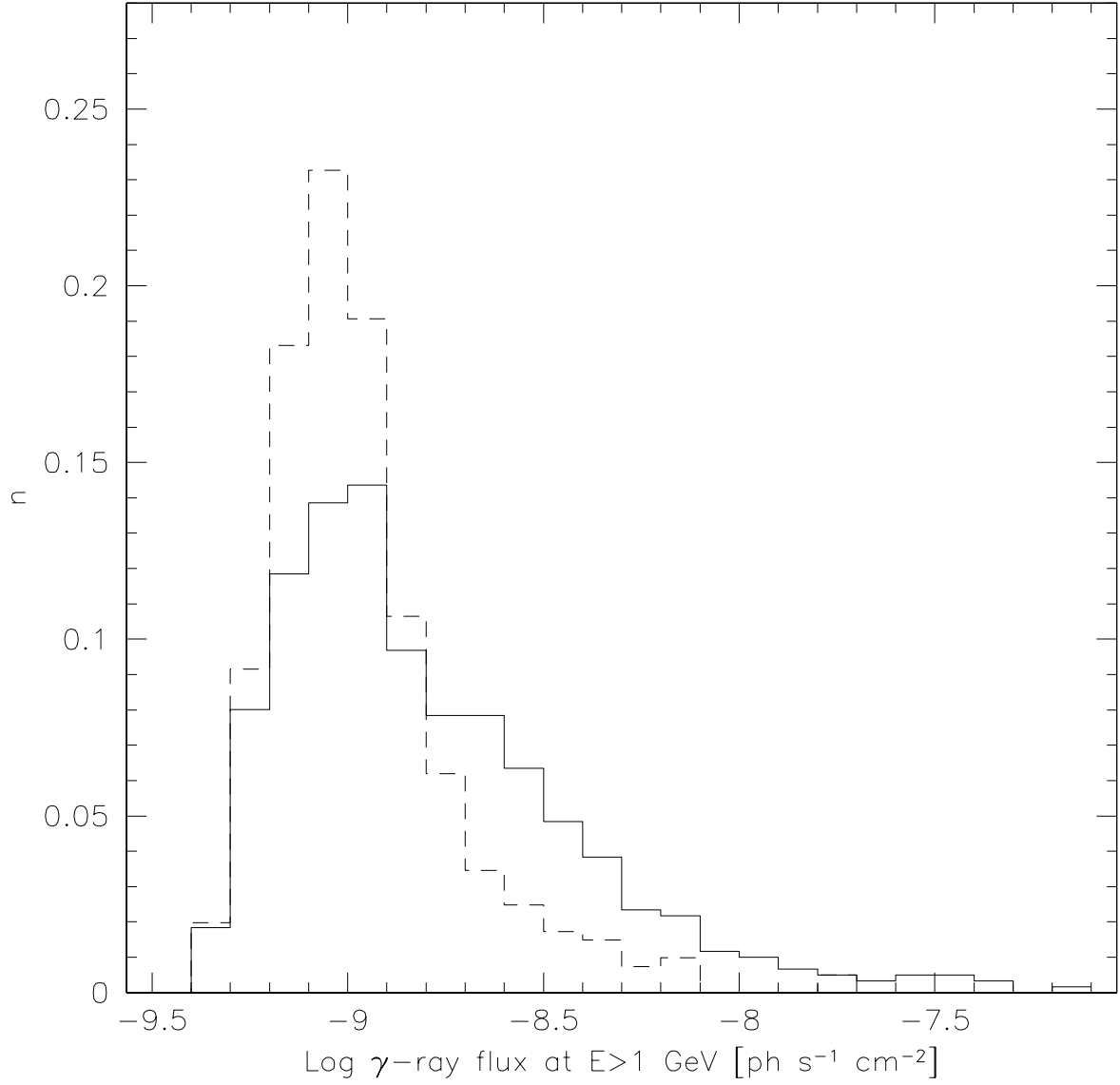}
\caption{Normalized distribution of the gamma-ray photon flux for high latitude ($|b| \ge 10^\circ$) associated (solid line) and unassociated (dashed line) 1FGL sources.  \label{f.gfluxhisto}}
\end{figure}

In our study we will of course be using only sources that have been associated
with a low-energy AGN counterpart. However, we note that 1FGL also contains 374
unassociated sources. If some of these sources are AGN that were not associated
with a lower-energy counterpart because they happen to be too faint in radio,
then this could potentially introduce a bias in our assessment of the
radio/gamma flux correlations. In Fig.~\ref{f.gfluxhisto}, we show normalized
histograms of the gamma-ray--fluxes of the high-latitude ($|b|>10^\circ$) AGNs
and of the high-latitude unassociated sources. Although in both distributions
the sources tend to cluster in the low flux bins, this effect is much more
pronounced in the unassociated gamma-ray sources, and there is strong
statistical evidence that the two samples are not drawn from the same
population (K-S probability of $4.3\times10^{-13}$).  This makes it unlikely
that we significantly overestimate the strength of the correlation because of
the existence of yet-unassociated, radio-faint and gamma-ray-bright blazars.

On the other hand, in any given radio flux limited sample there are sources
that are radio bright and gamma-ray quiet (see e.g.\ \S\ref{s.ovroobs} below
for the case of the OVRO sample). This fact can be the consequence of long-term
variability and/or low duty cycle in gamma-rays \citep{Ghirlanda2011}; in any
case, in this paper we only deal with the sources detected by LAT.

\subsection{Radio data}

In Table~\ref{t.radiodata}, we list the radio flux densities used for the
present work, along with some basic information on the sources (position,
optical and spectral type, redshift). In particular, we give the archival 8 GHz
interferometric flux density in Col.\,8 (with the corresponding reference in
Col.\,9) and the 15 GHz single dish flux density, when available, in
Cols.\,10--12. A summary of the details of the relevant observations are given
in the following subsections.

\subsubsection{CRATES/Other catalogs\label{s.crates}} 

For all sources in the 1LAC, we were able to collect interferometric
measurements of the historic radio flux density. This provides us with the
largest database of radio and gamma-ray measurements ever obtained and we use
it for a discussion of the correlation between the two bands.

Most of these data come from CRATES (478 sources) or CRATES-like (96 sources)
observations.  The CRATES catalog \citep{crates} contains precise positions,
8.4~GHz flux densities, and radio spectral indices for over 11,000
flat-spectrum sources over the entire $|b| > 10^\circ$ sky.  In the region
$\delta > -40^\circ$, the 8.4~GHz data were obtained with the VLA in its
largest (A) configuration, and the spectral indices were determined by
comparing the 8.4~GHz flux density and the 1.4~GHz flux density from the NRAO
VLA Sky Survey \citep[NVSS;][]{nvss}.  In the region $\delta < -40^\circ$, the
8.4~GHz data were obtained with ATCA in a variety of large configurations
(6A/C/D, 1.5B/C/D), and the spectral indices were determined by comparing the
8.4~GHz flux density and the 843~MHz flux density from the Sydney University
Molonglo Sky Survey \citep[SUMSS;][]{sumss}.\footnote{Strictly speaking, the
  ATCA observations were performed at 8.6~GHz, and the flux densities were
  converted to 8.4~GHz by interpolation using the spectral index, but even for
  a very inverted source ($\alpha = -1$), this represents an adjustment of
  $<$3\% to the flux density.}

The data for sources that are not in CRATES are often of identical or very
similar quality to those for CRATES sources.  For example, 8.4~GHz data from
the Cosmic Lens All-Sky Survey \citep[CLASS;][]{class1,class2}, from which the
CRATES catalog obtained much of its northern hemisphere data in the first
place, were all taken with the VLA in the A~configuration.  Similarly, the
PMN-CA catalog\footnote{Survey results can be downloaded from
  http://www.parkes.atnf.csiro.au/observing/databases/pmn/casouth.pdf} of over
6600 radio sources was compiled from 8.6~GHz data obtained with ATCA in the 6A,
6C, and 6D configurations.  As a result, the radio flux densities and spectral
indices of most non-CRATES sources can still be compared directly to those of
true CRATES sources without introducing any systematic errors or biases.

For 19 sources, for which 8.4 GHz VLA or ATCA measurements are not available,
we extrapolate from lower frequency interferometric measurements
\citep[e.g. those reported from the Roma-BZCAT,][]{bzcat}. The spectral indices
used for the extrapolation are those available from NED; when none was
available, it was conventionally set to $\alpha=0.0$.

Finally, there are 6 sources in the 1LAC that possess a significant amount of
extended radio emission \citep[such as the misaligned AGNs discussed
  by][]{MAGN} and escape the selection criteria of CRATES and similar
surveys. However, these are all rather well known radio sources, and it has
been straightforward to obtain interferometric measurements of their radio core
flux density from the literature, either directly or with trivial calculations
(e.g.\ interpolation).

\subsubsection{OVRO \label{s.ovroobs}}

Since late 2007, the Owens Valley Radio Observatory (OVRO) 40~m Telescope has
been engaged in a blazar monitoring program to support the \emph{Fermi}-LAT
\citep{OVRO}.  In this program, all 1158 CGRaBS blazars north of declination
$-20^\circ$ have been observed approximately twice per week or more frequently
since June~2007 \citep{cgrabs}.  Gamma-ray blazars and other sources detected
by {\em Fermi} have been added to the program which makes the total number of
monitored sources close to 1500. Of these sources, 199 appear as ``clean''
associations in the 1LAC catalog.

The OVRO flux densities are measured in a single 3~GHz wide band centered on
15~GHz.  Observations were performed using azimuth double switching as
described in \citet{Readhead1989}, which removes much atmospheric and ground
interference.  The relative uncertainties in flux density result from a 5~mJy
typical thermal uncertainty in quadrature with a 1.6\% systematic uncertainty.
The absolute flux density scale is calibrated to about 5\% via observations of
the steady calibrator 3C\,286, using the Baars et al.\ model
\citep{Baars1977}. A complete description of the OVRO program, population
studies of the radio variability, their relation with other physical properties
and a study of the time relation between radio and gamma-ray emission are
presented in a series of dedicated publications
\citep{OVRO,Methodology,MaxMoerbeckPrep}.

\begin{figure}
\plotone{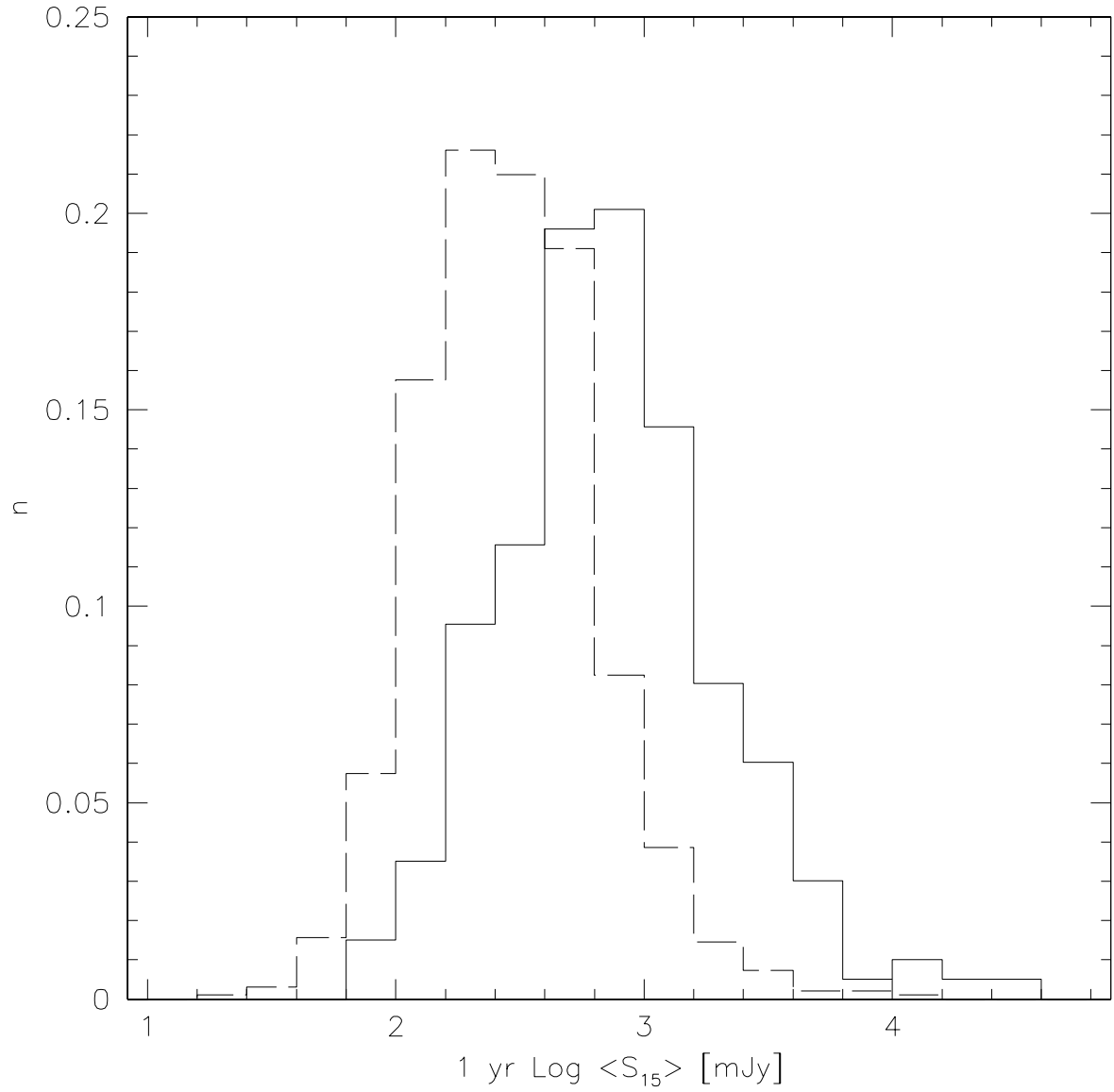}
\caption{Normalized distributions of the average 15 GHz flux densities for CGRABS sources north of declination $-20^\circ$, shown separately for gamma-ray associated (solid) and unassociated (dashed) sources. \label{f.ovro_flux_histo}}
\end{figure}

Because the \emph{Fermi}-LAT flux densities used in this study represent time
averages over the observation period, we produce estimates of the 15~GHz time
average flux density from the OVRO data for each source by linearly
interpolating between successive light curve values, integrating between the
start and end dates, then dividing by the time interval.  For the 11~month data
here, the start date was midnight August 4, 2008 (MJD 54682), and the end date
was midnight July 4, 2009 (MJD 55016).  Hereafter, we will be referring to
average 15 GHz radio fluxes obtained in this manner as the OVRO {\em
  concurrent} data.

qThe normalized distribution of average fluxes of the OVRO subset is shown in
Fig.~\ref{f.ovro_flux_histo}, over-plotted with the distribution of average
fluxes, obtained in the same manner, of gamma-ray quiet CGRaBS sources north of
declination $-20^\circ$. The sources which are also in 1LAC have generally
higher 15\,GHz average fluxes. However, there is substantial overlap between
the two distributions, so the existence of sources with large fluxes at 15 GHz
but which are faint in gamma rays is not unexpected. Therefore, our expectation
from the distribution of fluxes alone is that if a statistically significant
correlation between radio and gamma-ray fluxes indeed exists, it will likely
have a substantial scatter.

\section{Results \label{s.results}}

In this section we present the results of our search for possible correlations
between radio flux densities and the gamma-ray photon flux for the sources in
the 1LAC sample.  In particular, in Sect.~\ref{s.1lac} we consider the full
1LAC sample and search for correlations with archival radio data, while in
Sect.~\ref{s.ovro} we focus on the subset of sources observed at OVRO,
considering both concurrent and archival radio data; finally, in
Sect.~\ref{s.4bands} we present results for a subset of the 1LAC composed of
sources detected in at least 4 individual energy bands.  There are 599 sources
in the 1LAC clean sample and 199 in the 1LAC-OVRO sample.  The OVRO 15\,GHz
concurrent fluxes are averaged (time-integrated, as in the gamma-ray data) over
the same interval as the LAT observations, and for all of sources considered
here there exists {\em gamma-ray} variability on timescales shorter than the
averaging period.

For each sample, we have compared the radio flux density to the 1-yr gamma-ray
energy flux at $E>100$~MeV. Moreover, since we have unprecedentedly large
datasets, we can also explore whether the strengths of any observed
correlations are dependent on the gamma-ray energy band in which the flux is
calculated, or on the source spectral type.  For this reason, we also compare
radio flux densities to gamma-ray photon fluxes calculated in the single energy
bands $100\, \mbox{MeV} < E < 300\, \mbox{MeV}$, $300\, \mbox{MeV} < E < 1\,
\mbox{GeV}$, $1\, \mbox{GeV} < E < 3\, \mbox{GeV}$, $3\, \mbox{GeV} < E < 10\,
\mbox{GeV}$, $10\, \mbox{GeV} < E < 100\, \mbox{GeV}$. In each energy band, we
consider only the sources that are significant in that band. Not every source
is detected in all energy bands; actually, only a small minority is,
i.e. 51/599 (8.5\%). As a consequence of their different spectral properties,
FSRQs are generally more abundant in the lowest energy bands, while BL Lacs are
more numerous in the most energetic ones. For instance, for the 1LAC sources,
we have 128 FSRQs and 47 BL Lacs in the 100--300 MeV band, and 22 FSRQs and 99
BL Lacs in the 10-100 GeV band.

Since FSRQs and BL Lacs have different spectral properties and showed different
behaviors in the preliminary analysis \citep{LBAS,Giro10}, we also tested the
two populations separately, in addition to the full set of sources. Moreover, a
classification based on the broadband spectral properties is physically more
meaningful, so we also consider the populations of low-, intermediate-, and
high-synchrotron-peaked blazars (LSP, ISP, and HSP respectively).

In total, we have 36 combinations of source type and gamma-ray energy band for
the 1LAC. For the OVRO sample, we have also the possibility to consider the
radio data obtained at 15 GHz during the same interval of the gamma-ray
observations, both as mean and peak flux density measurements, and in a
different time domain. For each combination, we produced a scatter plot of the
radio vs.\ gamma-ray flux densities and determined the Spearman's rank
correlation $\rho$, which are presented in the following subsections.

The value of $\rho$ is characteristic of the strength of the correlation, and
it can be related to the significance of an {\it apparent} correlation between
radio and gamma-ray fluxes. However, an assessment of the statistical
significance of an {\it intrinsic} correlation in each case (after the effects
of a common distance and a limited dynamical range are accounted for) is
nontrivial and cannot be based on a conventional assumption of unbiased
samples. Therefore, we use simulations based on the method of surrogated data
to evaluate the significance of intrinsic correlations and discuss them in
Sect.~\ref{s.mc}.

\subsection{Full sample \label{s.1lac}}

The sources associated to the 1LAC members span over 4 orders of magnitude in
radio flux density, ranging between a few mJy for the faintest BL Lacs to
several 10's of Jy for the brightest quasars (e.g.\ 3C\,273 and 3C\,279). The
flux density distributions for the whole population and divided by source type
are shown in \citet{1LAC}. The overall distribution shows a broad peak at
$S\sim 800$ mJy, which is the result of the combination of the two peaks of the
single population distributions, with BL Lacs peaking around $S \sim 400$ mJy
and FSRQs at $S \sim 1300$ mJy. In gamma-rays, the energy fluxes span over 3
orders of magnitude (between $4.8 \times 10^{-12}$ and $6.6 \times 10^{-10}$
{\mbox{${\rm \, erg \,\, cm^{-2} \, s^{-1}}$}} at $E>100$\,MeV), with BL Lacs
typically fainter than FSRQs; the mean photon fluxes at $E>100$ MeV are $8.5
\times 10^{-8}$ \pflux{} and $2.9 \times 10^{-8}$ \pflux{} for FSRQs and
BL~Lacs respectively \citep{1LAC}.

\begin{figure}
\plotone{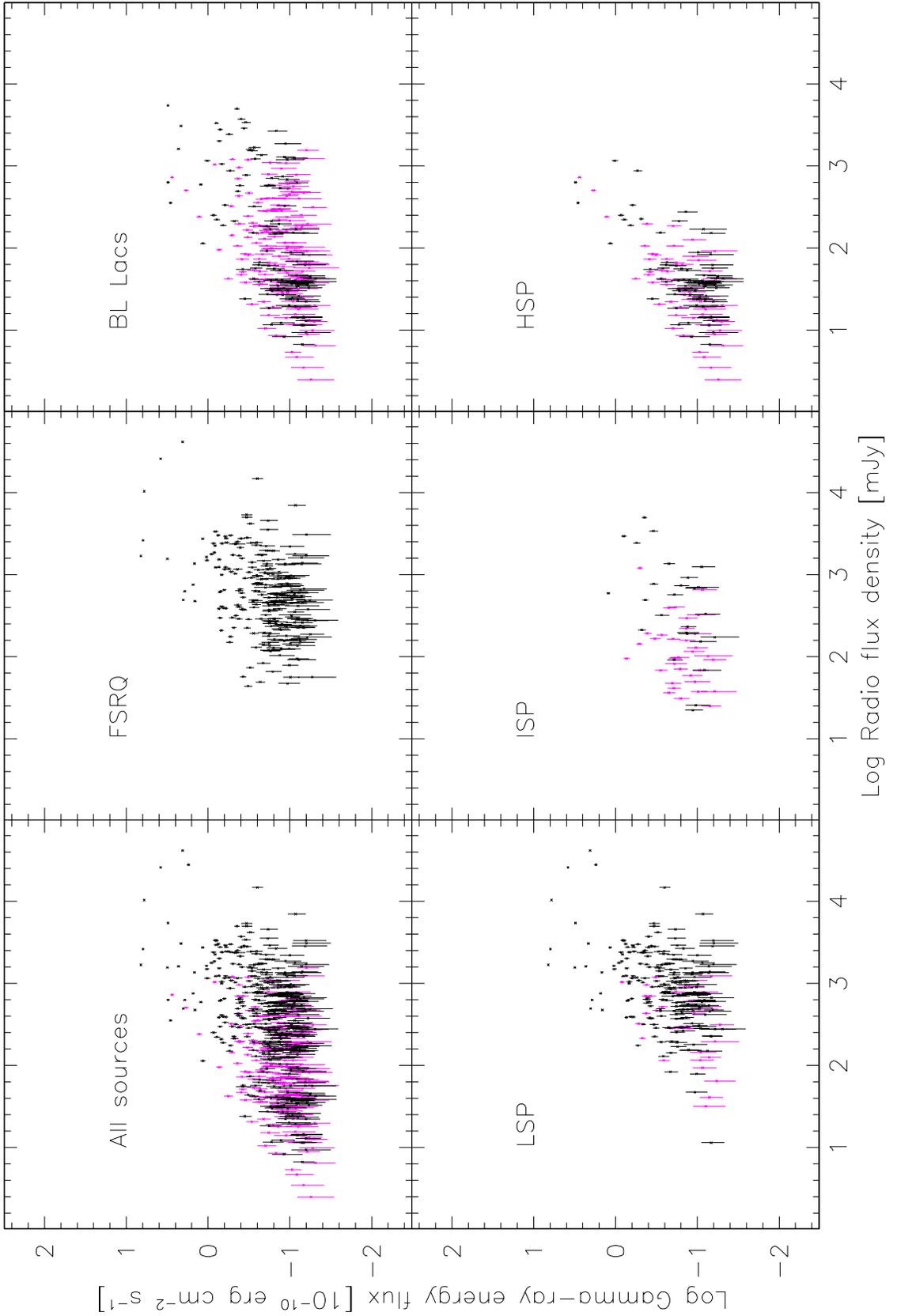}
\caption{Broad band gamma-ray energy flux vs.\ 8 GHz archival radio flux density for the
  1LAC sample, divided by source type. Top, from right to left:\ all AGNs, FSRQ, BL Lacs; bottom, from right to left:\ LSP, ISP, and HSP blazars. Sources with unknown redshift are shown in magenta. \label{f.1lac-sum}}
\end{figure}

\begin{figure}
\plotone{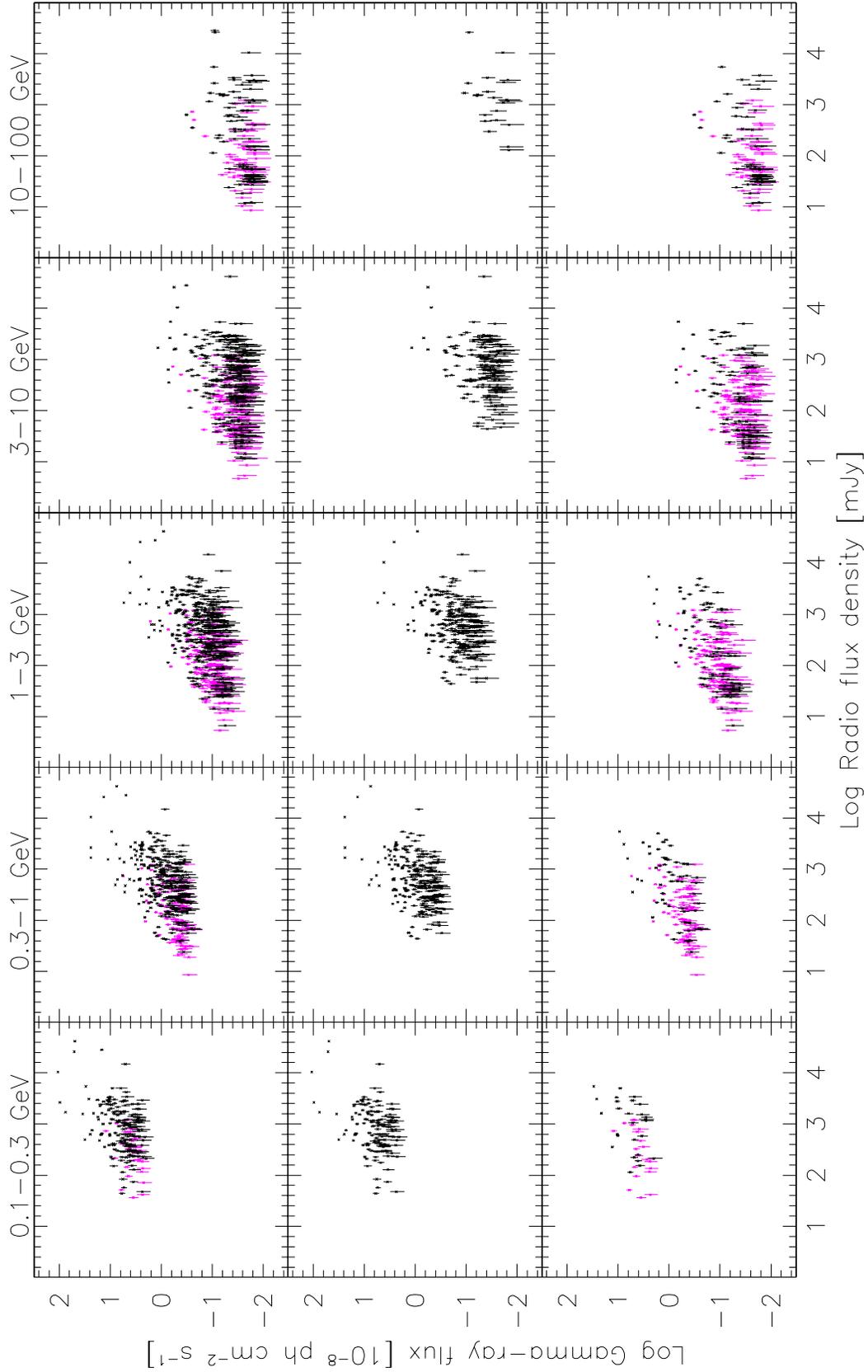}
\caption{Gamma-ray photon flux vs.\ 8 GHz archival radio flux density for the
  1LAC sample, divided by source optical type (top:\ all sources, middle:\ FSRQ, bottom:\ BL Lacs) and in energy bands. Sources with unknown redshift are shown in magenta. \label{f.1lac-opt}}
\end{figure}

\begin{figure}
\plotone{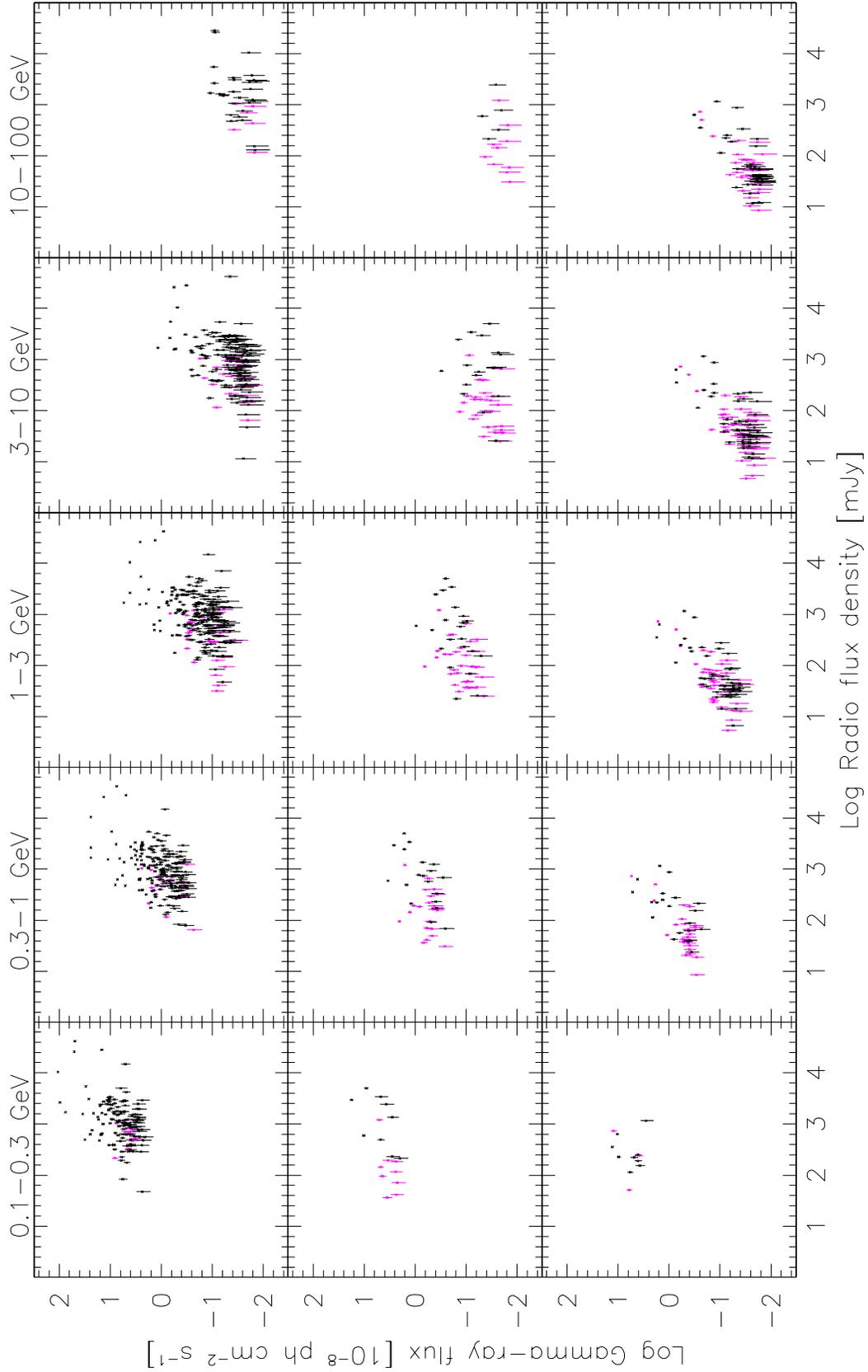}
\caption{Gamma-ray photon flux vs.\ 8 GHz archival radio flux density for the
  1LAC sample, divided by source spectral type (top:\ LSP, middle:\ ISP,
  bottom:\ HSP) and in energy bands. Sources with unknown redshift are shown in
  magenta. \label{f.1lac-sed}}
\end{figure}

We show the gamma-ray and radio flux scatter plots for the 1LAC sources in
Fig.~\ref{f.1lac-sum}, \ref{f.1lac-opt}, and \ref{f.1lac-sed}.  Each figure
shows a collection of panels showing various combinations of the 1FGL gamma-ray
flux and radio historical flux density. In particular, Fig.~\ref{f.1lac-sum}
shows the gamma-ray energy flux vs.\ radio flux density for all sources (top
left panel), sources divided by optical type (FSRQ and BL Lacs in the centre
and right top panels, respectively), and sources divided by spectral type
(bottom row, with LSP, ISP, and HSP in the left, middle, and right panels,
respectively); in Fig.~\ref{f.1lac-opt} and Fig.~\ref{f.1lac-sed}, we show the
gamma-ray photon flux vs.\ radio flux in the five individual LAT energy bands
(left to right), divided by source type:\ in Fig.~\ref{f.1lac-opt}, the top row
shows all sources, the middle one shows FSRQs, and the bottom one BL Lacs; in
Fig.~\ref{f.1lac-sed}, top, middle, and bottom rows are for the three different
synchrotron peak classes: LSP, ISP, and HSP blazars, respectively. Symbols in
magenta show sources for which a redshift is not available.

\begin{figure}
\plotone{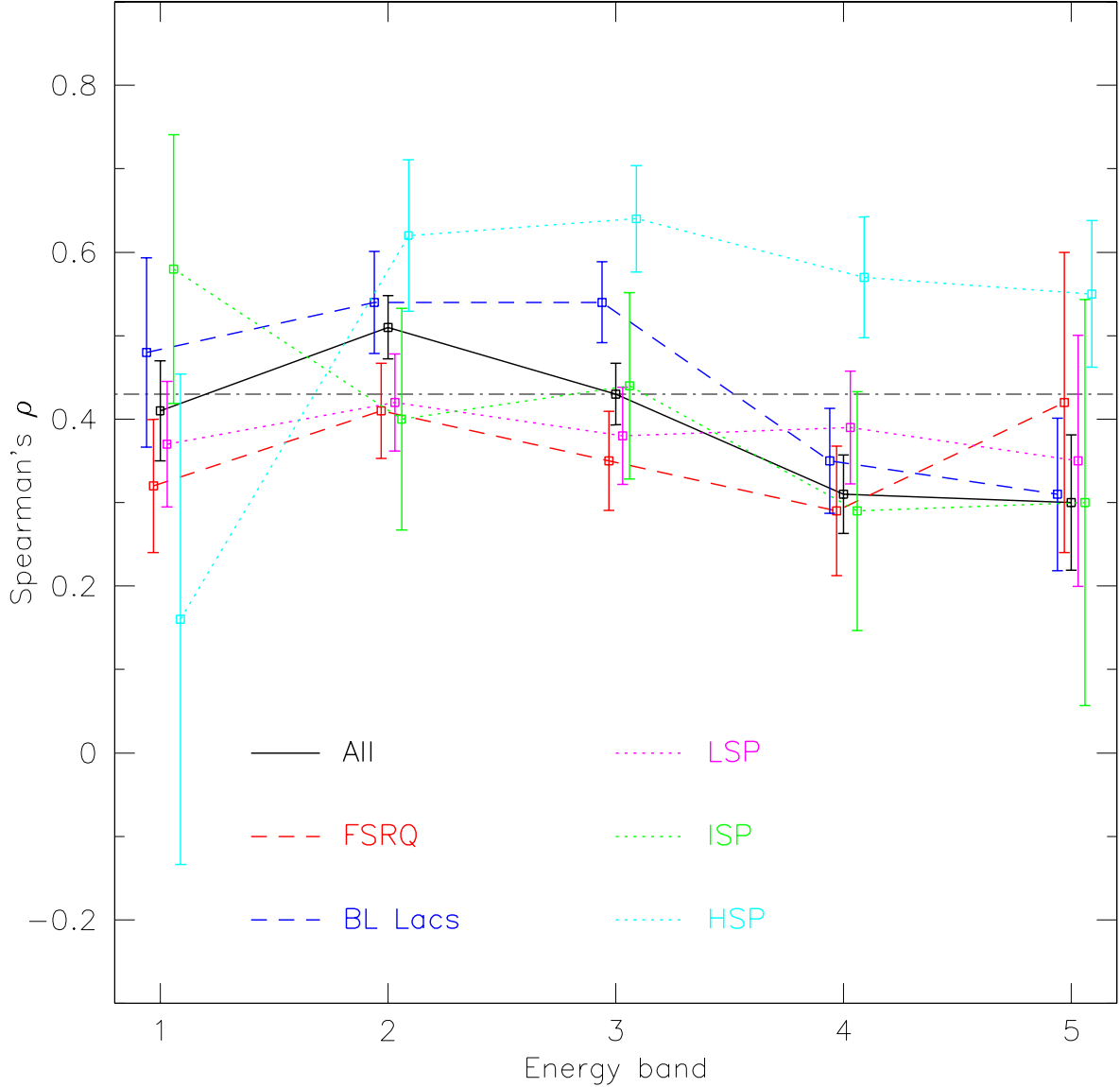}
\caption{Correlation coefficient for the 1LAC sample as a function of the energy bands. Solid black line: all sources; dashed lines: sources divided by optical type: red for FSRQs and blue for BL Lacs; dotted lines: sources divided by spectral type (LSP  in magenta, ISP in green, HSP in cyan). The dot-dash black line shows as a reference the value of $\rho$ obtained using all sources and broad band gamma-ray flux. At each $x$-point (energy band), symbols are horizontally offset for improved clarity. \label{f.1lac-rho}}
\end{figure}

We report the correlation coefficients between radio and gamma-ray flux for the
full sample in Table \ref{t.1lac}, divided by source type and energy band, and
we visualize them in Fig.~\ref{f.1lac-rho}. In this figure, the correlation
coefficients are shown across the five energy bands and are connected with
lines of different color and style for the various sub-populations:\ solid
black line for the full 1LAC sample, dashed lines for optical type sub-groups
(red for FSRQ and blue for BL Lacs), dotted lines for sub-groups defined by the
spectral properties (magenta for LSP, green for ISP, cyan for HSP). The
accuracy to which the correlation coefficients are determined, based on the
number of sources and the strength of the correlation, is shown by the error
bars, which correspond to the standard deviation for $\rho$, defined as
$\sigma_\rho = (1-\rho^2) / \sqrt{N-1}$; although this standard deviation is
formally defined only for the Pearson product-moment correlation coefficient
$r$ \citep{Wall2003}, we extend it to our case, since the distribution of the
Spearman $\rho$ for $N>30$ approaches that of the Pearson product-moment.

The Spearman correlation coefficient for all (599) sources is $\rho=0.43$.
FSRQs and BL Lacs reveal different behaviors.  In general, BL Lacs exhibit
larger values of $\rho$ than FSRQs, both when the broad band gamma-ray energy
flux is considered and in most of the single energy bands; for example, in the
most populated energy band (for both populations, the 1--3 GeV band, with 220
FSRQ and 214 BL Lacs), we find $\rho=0.54$ for BL Lacs and $\rho=0.35$ for
FSRQs, although the difference is less significant in the other energy
bands. Moreover, in FSRQs the correlation coefficient is quite stable across
the various energy ranges (between $\rho=0.29$ and $\rho=0.42$), while BL Lacs
display some evolution, with $\rho$ decreasing as fluxes at higher energy bands
are considered.  If one looks at the spectral type populations, HSPs are always
the ones showing a tighter apparent correlation (except for the scarcely
populated 100--300 MeV band), and as high as $\rho=0.64$ in the 1--3 GeV band.

\subsection{OVRO sample \label{s.ovro}}

The sources with OVRO data represent a 199 element subset of the 1LAC sample,
going down to radio fluxes as low as 172 mJy (archival 8\,GHz value for
J1330+5202, the source associated to 1FGL\,J1331.0+5202) and 64.7 mJy (1-yr
concurrent 15 GHz value for J1725+1152). FSRQs outnumber BL Lacs by 120/69.
This sample provides the largest dataset of concurrent radio measurements to
the 1LAC fluxes and is therefore highly valuable in order to understand the
implications of variability on the radio/gamma-ray correlation.

In particular, we are in the position of comparing the correlation coefficient
not only among different source types and energy bands, but also to assess the
differences that arise when we use concurrent data or not. In
Table~\ref{t.ovro}, we give the correlation coefficients: for the
radio/gamma-ray flux densities using historical radio flux densities at 8 GHz;
the mean and peak flux density value at 15 GHz calculated over the first 11
months of activity of the LAT; and an average 15 GHz flux calculated over a
one-week interval {\em after} the first 11 months of activity of the LAT
(specifically, the period between January 23 -- 31 2010).

\begin{figure}
\plotone{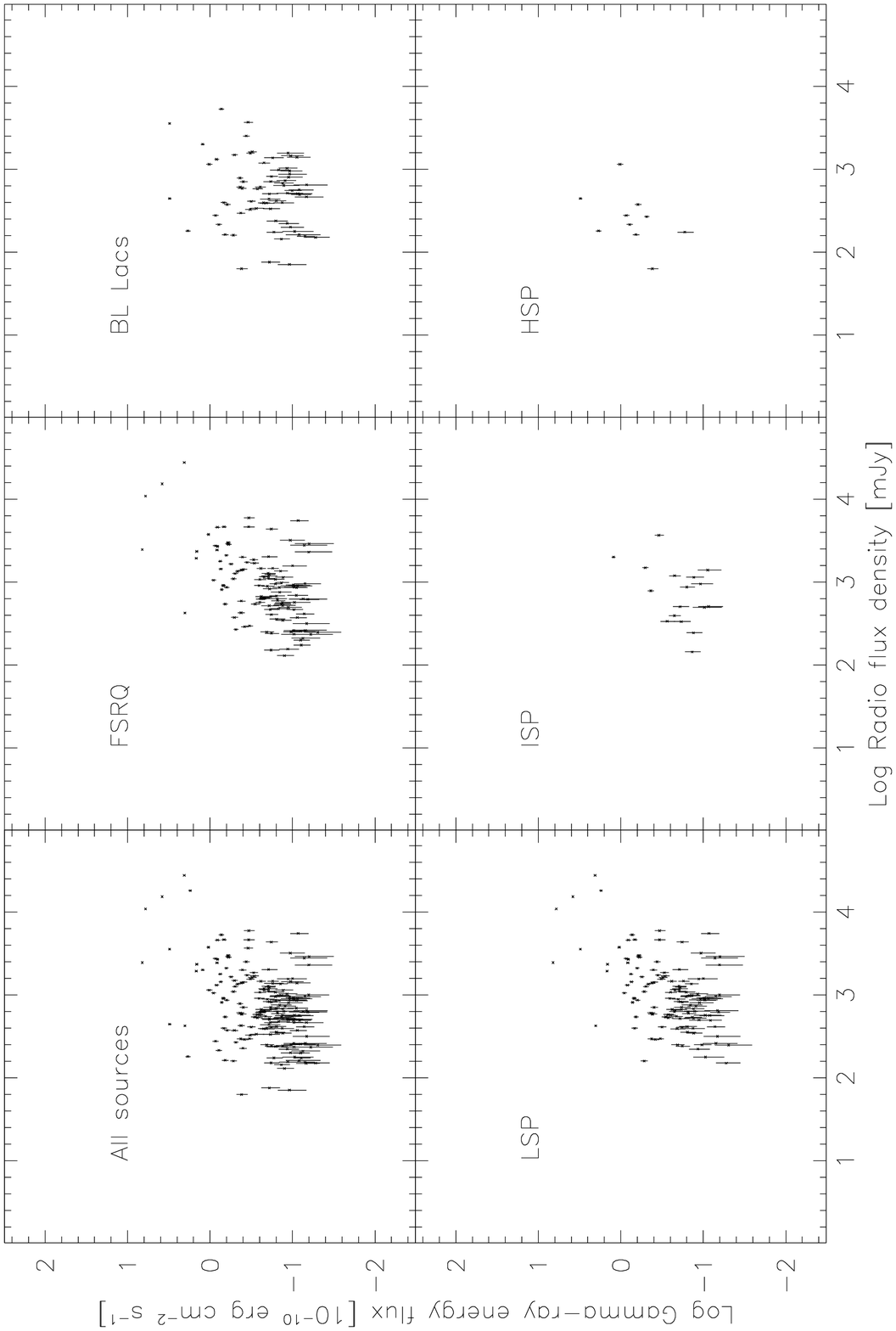}
\caption{Broad band gamma-ray energy flux vs.\ concurren 15 GHz mean radio flux density for OVRO sources, divided by source type. Top, from right to left:\ all AGNs, FSRQ, BL Lacs; bottom, from right to left:\ LSP, ISP, and HSP blazars. Sources with unknown redshift are shown in magenta. \label{f.ovro15-sum}}
\end{figure}

\begin{figure}
\plotone{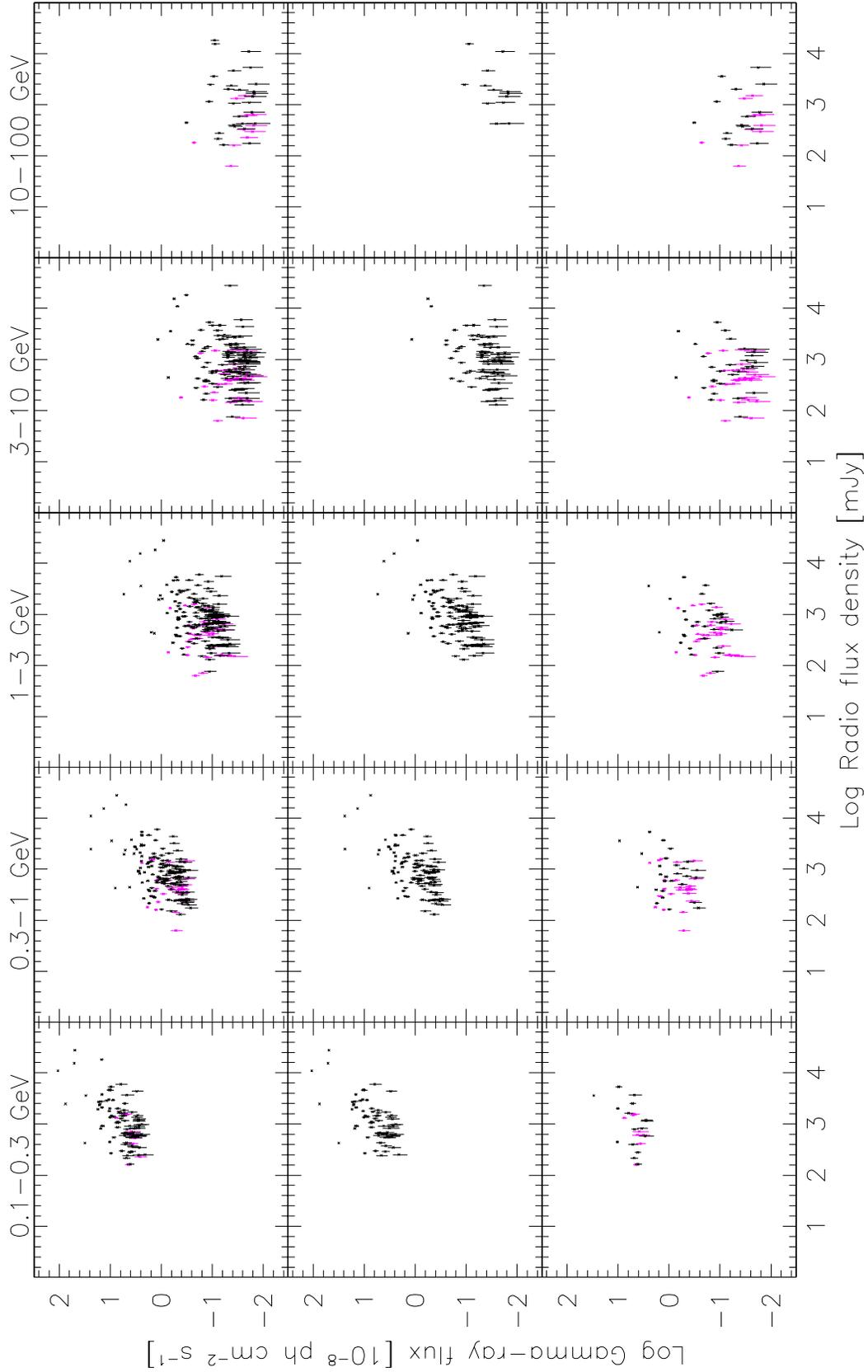}
\caption{Gamma-ray photon flux vs.\ concurrent 15 GHz mean radio flux density
  for OVRO sources, divided by source optical type (top:\ all sources,
  middle:\ FSRQ, bottom:\ BL Lacs) and in energy bands. Sources with unknown
  redshift are shown in magenta. \label{f.ovro15-opt}}
\end{figure}

\begin{figure}
\plotone{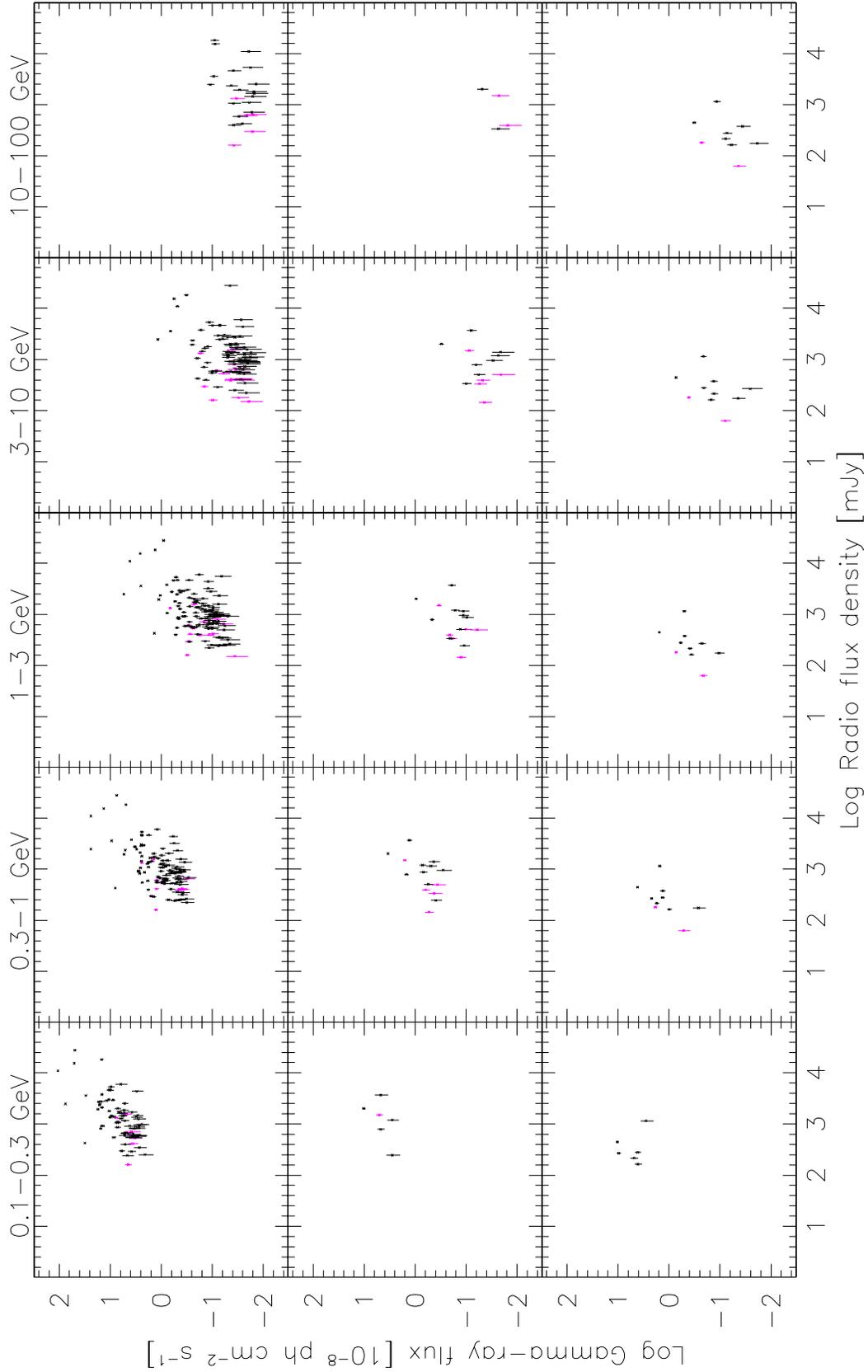}
\caption{Gamma-ray photon flux vs.\ concurrent 15 GHz mean radio flux density
  for OVRO sources, divided by source spectral type (top:\ LSP, middle:\ ISP,
  bottom:\ HSP) and in energy bands. Sources with unknown redshift are shown in
  magenta. \label{f.ovro15-sed}}
\end{figure}

\begin{figure}
\plotone{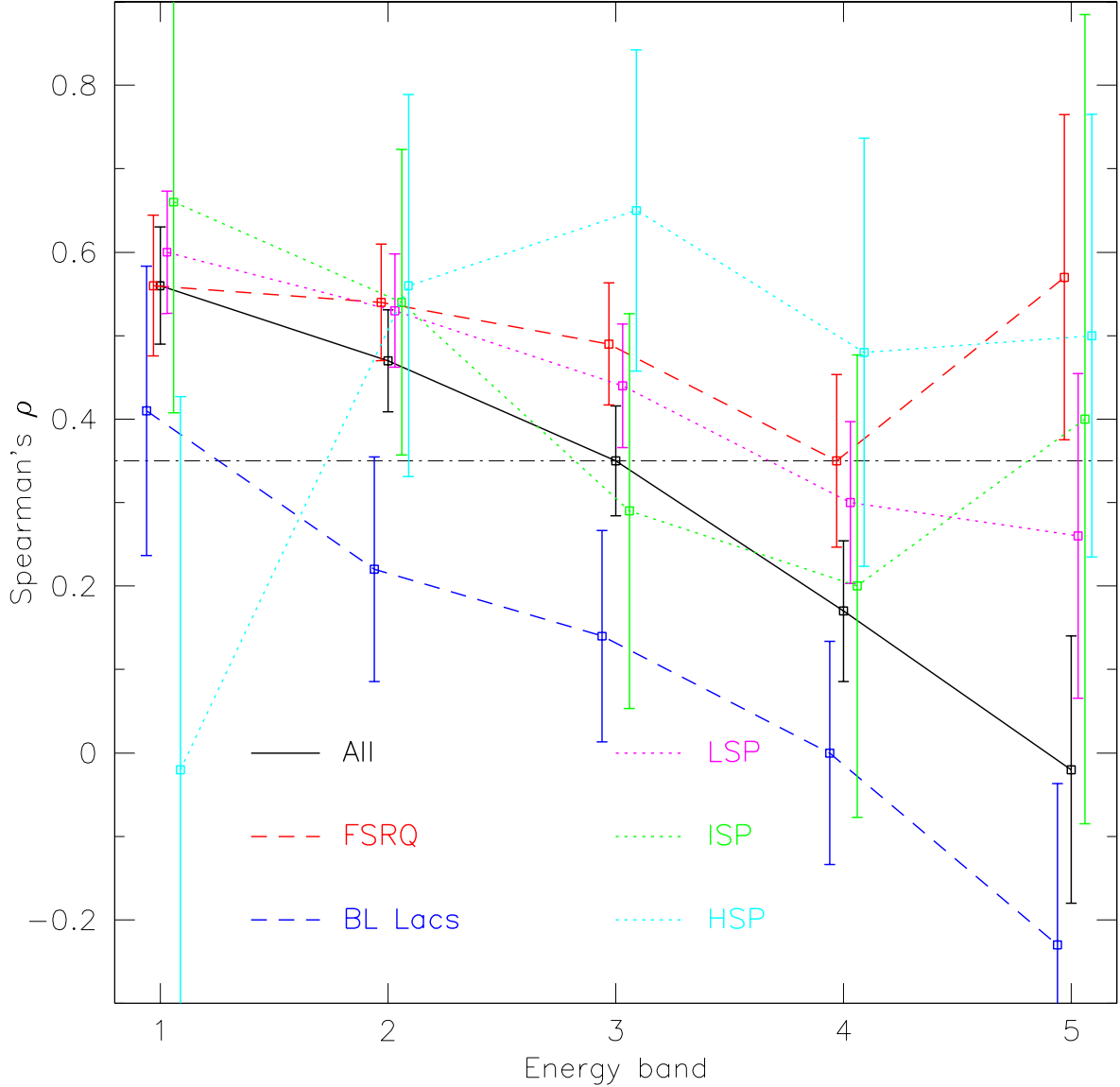}
\caption{Correlation coefficients for the OVRO sample (concurrent radio and gamma-ray data) as a function of energy bands. Solid black line: all sources; dashed lines: sources divided by optical type: red for FSRQs and blue for BL Lacs; dotted lines: sources divided by spectral type (LSP in magenta, ISP in green, HSP in cyan). The dot-dash black line shows as a reference the value of $\rho$ obtained using all sources and broad band gamma-ray flux. At each $x$-point (energy band), symbols are horizontally offset for improved clarity.  \label{f.ovro15-rho}}
\end{figure}

Figures~\ref{f.ovro15-sum}, \ref{f.ovro15-opt} and \ref{f.ovro15-sed} show the
scatter plots of the concurrent radio and gamma-ray fluxes, using mean values
for the radio flux density. As for the 1LAC case, we show 3 collections of
scatter plots: radio vs.\ gamma-ray energy flux for all sources, FSRQ, BL Lacs,
LSP, ISP, and HSP sources in Fig.~\ref{f.ovro15-sum}, radio vs.\ gamma-ray
photon flux for all sources, FSRQ, and BL Lacs in Fig.~\ref{f.ovro15-opt}, and
for LSP, ISP, and HSP sources in Fig.~\ref{f.ovro15-sed}. Finally, the trend of
$\rho$ as a function of energy band for the various sub-classes is shown in
Fig.~\ref{f.ovro15-rho}, with the same notation as in Fig.~\ref{f.1lac-rho}.

Unlike for the larger 1LAC sample, in the sample with concurrently-measured
radio fluxes FSRQs generally display larger values of $\rho$ than BL Lacs; as
an example, in the 1--3~GeV energy band, $\rho_\mathrm{FSRQ}=0.48$ and
$\rho_\mathrm{BLL} =0.13$. Moreover, the correlation coefficient for BL Lacs
for the energy bands above 1 GeV is consistent with no correlation, becoming
even marginally negative in the 10--100 GeV band. It has to be remembered that
the OVRO sample is somewhat biased in favor of bright radio sources, so it
contains relatively few BL Lacs, and in particular just a handful (10/199) of
HSPs, as they are generally rather radio weak.  Interestingly, the BL Lac curve
falls below the three individual spectral type curves (LSP, ISP, HSP). We note
that the sample also contains 41 sources (about 20\% of the total) whose
spectral type is unknown and that have almost uncorrelated radio and gamma-ray
flux density. While this explains part of the difference between BL Lacs and
LSP+ISP+HSP, it is also important to warn that the 33 LSP BL Lacs do show
systematically lower values of $\rho$ than the whole group of LSP sources
(which is dominated by FSRQs).

As far as the radio variability is concerned, we find that for the whole sample
the correlation coefficient with concurrent 15 GHz data is always larger than
that obtained using archival 8 GHz data or non-concurrent 15 GHz OVRO
data. This result is mostly driven by the FSRQ population, while the less
numerous BL Lac population does not seem to reveal significant differences
between the use of concurrent or non-concurrent radio data.  Finally, the use
of the peak 15 GHz flux density yields generally weaker correlations, in some
cases even weaker than those found using non-concurrent data.

\subsection{Sources significant in at least four energy bands \label{s.4bands}}

For both the 1LAC and the OVRO samples, we have considered in each energy range
all the sources that were significant in that band. As a consequence of the
different spectral characteristics of each individual source, the samples used
to calculate the various coefficients have often little overlap between each
other (even within the same population), particularly in energy bands that are
far apart.

For this reason, we have also considered a third case, the sample of sources
that are significant in at least four of the five individual energy bands.  In
this way, we build a relatively bright, well defined, and sizable sample. This
sample is composed of 192 sources, and both FSRQ (94 sources) and BL Lacs (84)
are well represented.

\begin{figure}
\plotone{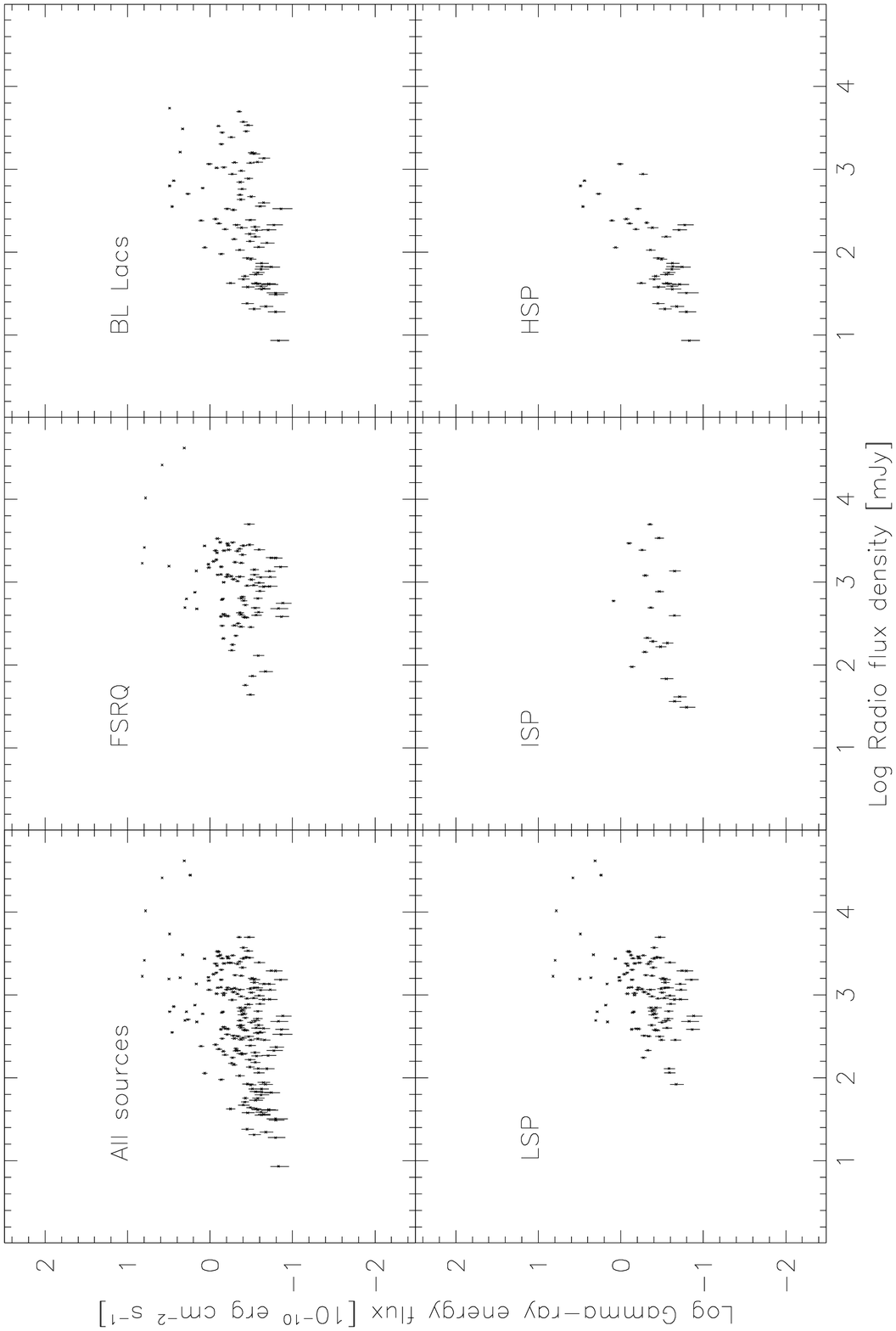}
\caption{Broad band gamma-ray energy flux vs.\ 8 GHz archival radio flux density for sources in the
  1LAC with a detection in at least 4 energy bands, divided by source type. Top, from right to left:\ all AGNs, FSRQ, BL Lacs; bottom, from right to left:\ LSP, ISP, and HSP blazars. Sources with unknown redshift are shown in magenta. \label{f.4bands-sum}}
\end{figure}

\begin{figure}
\plotone{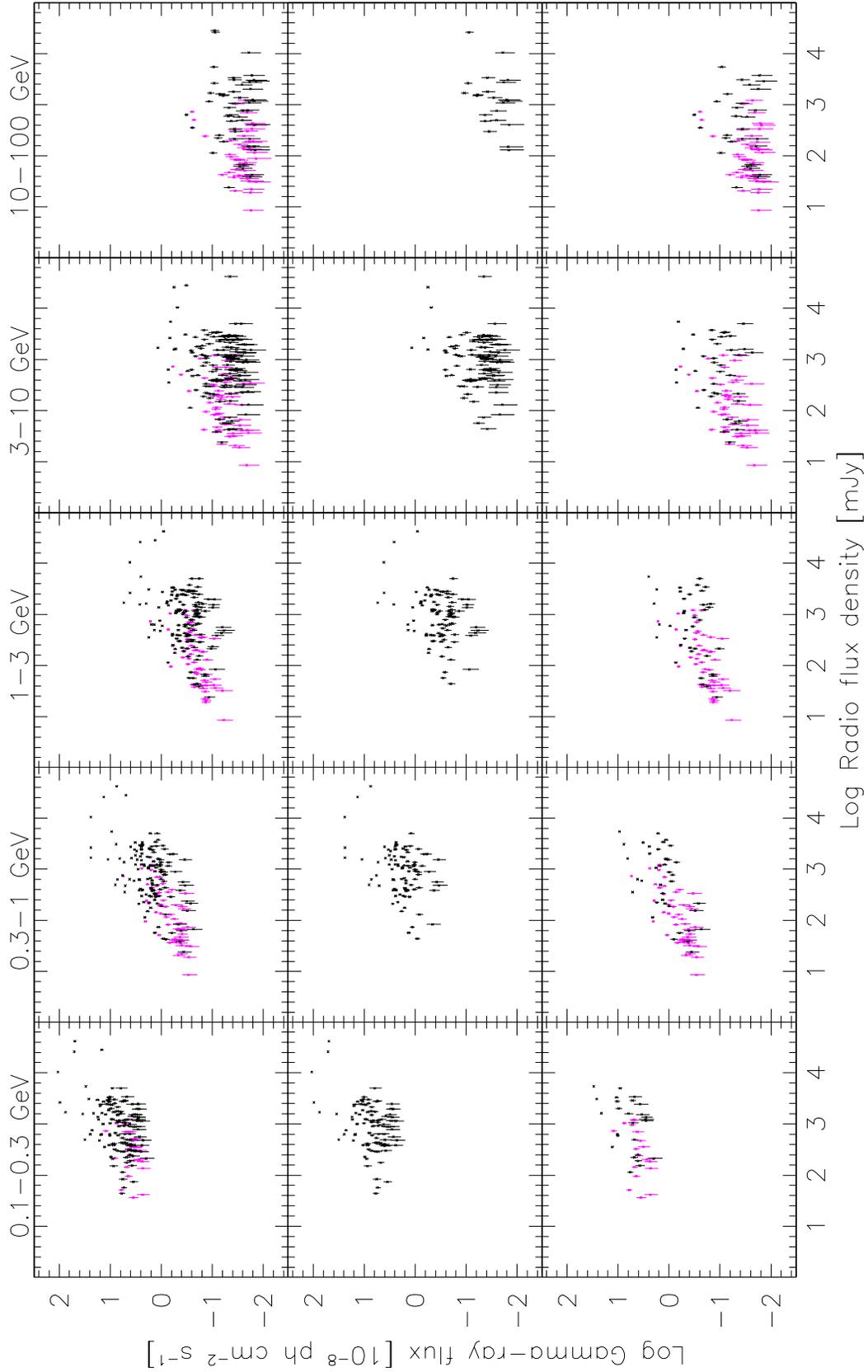}
\caption{Gamma-ray photon flux vs.\ 8 GHz radio flux density for sources
  in the 1LAC with a detection in at least 4 energy bands, divided by source
  optical type (top:\ all sources, middle:\ FSRQ, bottom:\ BL Lacs) and in
  energy bands. Sources with unknown redshift are shown in
  magenta. \label{f.4bands-opt}}
\end{figure}

\begin{figure}
\plotone{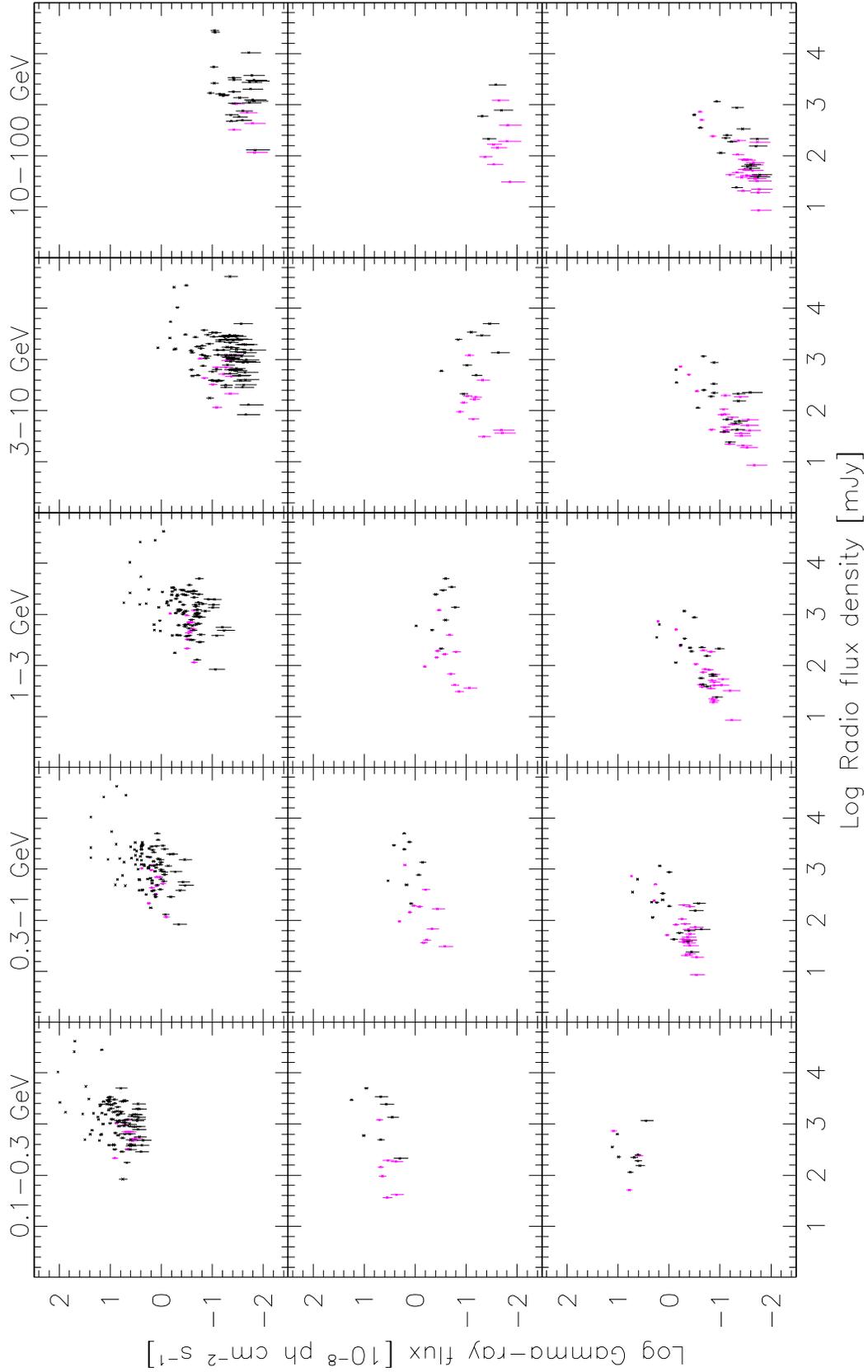}
\caption{Gamma-ray photon flux vs.\ 8 GHz radio flux density for sources
  in the 1LAC with a detection in at least 4 energy bands, divided by source
  spectral type (top:\ LSP, middle:\ ISP, bottom:\ HSP) and in energy
  bands. Sources with unknown redshift are shown in
  magenta. \label{f.4bands-sed}}
\end{figure}

\begin{figure}
\plotone{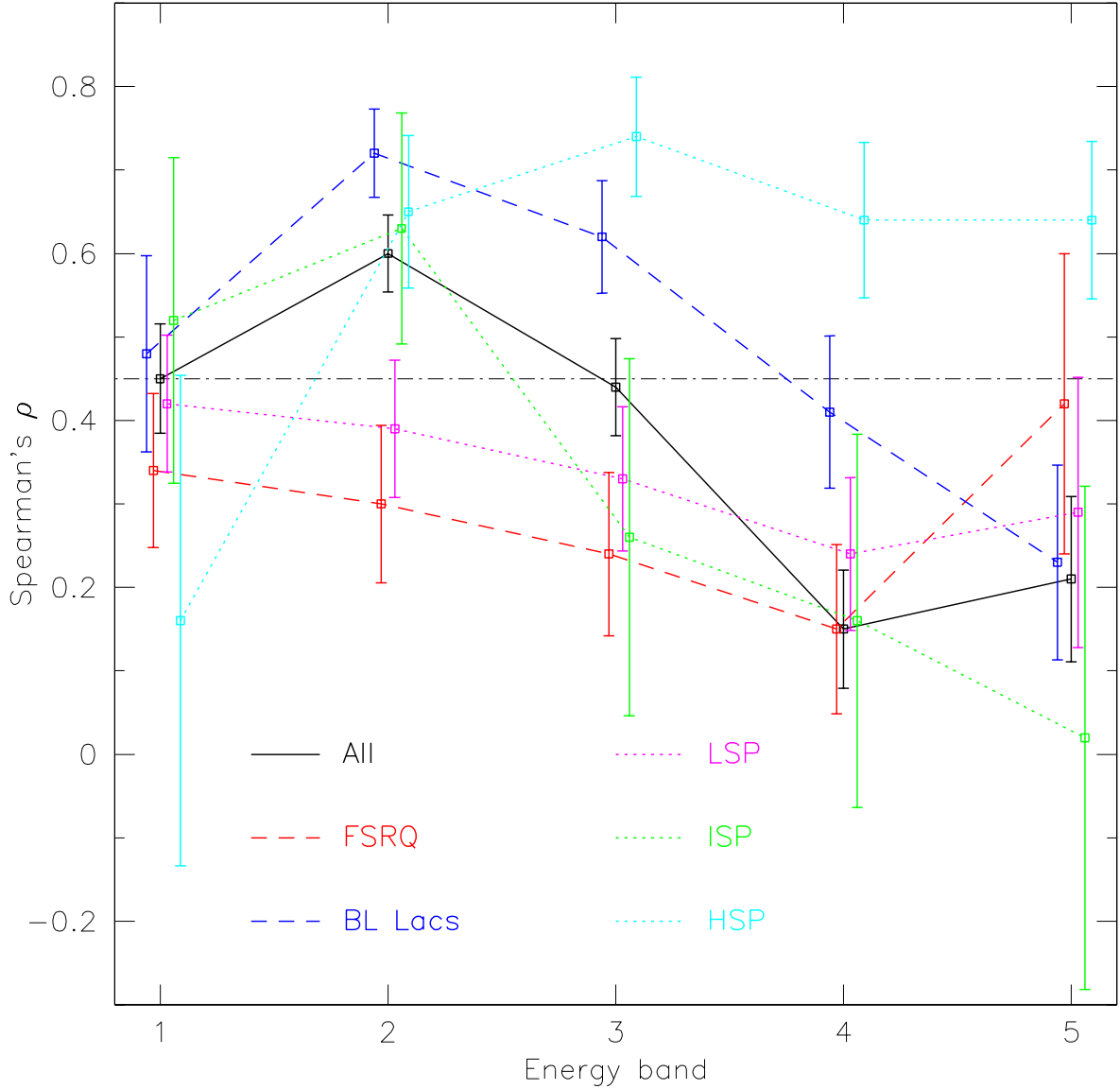}
\caption{Correlation coefficient for sources in the 1LAC with a detection in
  at least 4 energy bands, as a function of the energy bands. Solid black
  line: all sources; dashed lines: sources divided by optical type: red for
  FSRQs and blue for BL Lacs; dotted lines: sources divided by spectral type
  (LSP in magenta, ISP in green, HSP in cyan). The dot-dash black line shows as a reference the value of $\rho$ obtained using all sources and broad band gamma-ray flux. At each $x$-point (energy band), symbols are horizontally offset for improved clarity. \label{f.4bands-rho}}
\end{figure}

As in the full 1LAC sample, BL Lacs have generally higher values of $\rho$ than
FSRQ (e.g., $\rho_\mathrm{BLL}=0.54$ and $\rho_\mathrm{FSRQ} =0.29$ for the
radio vs.\ energy flux at $E>100$~MeV correlation). The individual values are
reported in Table~\ref{t.4bands} and Figs.~\ref{f.4bands-sum},
\ref{f.4bands-opt}, \ref{f.4bands-sed}, and \ref{f.4bands-rho}, using the same
notation as in the full 1LAC and OVRO cases.

If we look at the three groups defined by the synchrotron spectral properties,
we find that the maximum of the correlation coefficient is obtained in the
lowest energy band for LSP ($\rho=0.41$ between 100 and 300 MeV), in the 300
MeV--1 GeV for ISP ($\rho=0.63$), and in the 1--3 GeV for HSP
($\rho=0.74$). Therefore -- albeit with some overlap between the error bars --
the higher the spectral frequency of the synchrotron spectral peak, the higher
the energy at which the strongest apparent correlation is observed, and the
higher the correlation coefficient itself.

\section{Significance tests with the method of surrogate data \label{s.mc}}

In order to quantitatively assess the significance of any apparent correlation
between concurrent radio and gamma-ray flux densities of blazars in the
presence of distance effects, we have used a test based on the method of
surrogate data. In studying possible intrinsic correlations between flux
densities in different bands the null hypothesis is that they are {\em
  intrinsically} uncorrelated (implicitly assuming that any apparent
correlation is due to the observational errors and/or biases).  In a
frequentist approach, we investigate how frequently a sample of objects with
intrinsically uncorrelated gamma/radio flux densities, similar to the sample at
hand, will yield an apparent correlation as strong as the one seen in the data,
when subjected to the same distance effects as our actual sample
\citep[see][for a more detailed description of the test]{Methodology}.

In our test the strength of the apparent correlation is quantified by the Pearson product-moment correlation coefficient $r$, defined as
\begin{equation} \label{rcoeff}
r = \frac{\sum_{i=1}^N(X_i-\bar{X})(Y_i-\bar{Y})}
{\sqrt{\sum_{i=1}^N(X_i-\bar{X})^2\sum_{i=1}^N(Y_i-\bar{Y})^2}}
\end{equation}

Since it is not always straightforward to construct simulated samples with the
exact same selection criteria as the data sample, we have used {\em
  permutations} of measured quantities. To simulate the effect of a common
distance on intrinsically uncorrelated luminosities, we permute in luminosity
space:

\begin{itemize}
\item We split our sample in N redshift bins, with N determined so that each bin has at least $\sim 10$ sources. The separation in bins ensures that the luminosity and redshift distributions of the simulated samples approximate those in the real data, thus avoiding the introduction of biases not present in the data. Note however that, as we have shown in detail in \citet{Methodology}, the significance of the correlation we find increases with increasing N (the correlation becomes more significant), until it saturates for large enough N, provided that the number of sources is large enough. 
\item In each redshift bin: from the measured radio and gamma-ray flux densities, we calculate radio and gamma-ray luminosities at a common rest-frame radio frequency and rest-frame gamma-ray energy.\footnote{In order to implement the K-correction (project our calculated luminosities to a common rest-frame frequency in each band) we are using the historical radio spectral index $\alpha$ and the 1FGL photon index; The spectral index has been shown, at least at radio frequencies, to vary with flux \citep{Bonn2010};  however, as shown in \citet{Methodology}, different choices in radio spectral indices do not have a large effect on the resulting correlation significance, as the sources of interest generally have flat radio spectra and the relevant K-correction is small.}
\item We permute the evaluated luminosities, to simulate objects with {\em intrinsically uncorrelated} radio/gamma luminosities.
\item We assign a common redshift (one of the redshifts of the objects in the bin, randomly selected) to each luminosity pair, and return to flux-density space. Returning to flux-density space allows us to avoid Malmquist bias; assigning a common redshift allows us to simulate the common-distance effect on uncorrelated luminosities. In addition, by permuting in luminosity space we are guaranteed that the simulated samples have the same luminosity dynamical range as our actual sample.
\item To avoid apparent correlations induced by a single very bright or very faint object {\em much brighter or fainter than the objects in our actual sample}, we reject any flux-density pairs where one of the flux densities is outside the flux-density dynamical range in our original sample. The rejection rate is however very low for N$\geq 3$, and it decreases with increasing N. 
\end{itemize}

Using a number of flux density pairs equal to the number of objects in our
actual sample, we calculate a value for $r$. We repeat the process a large
number of times, and calculate a distribution of $r-$values for intrinsically
uncorrelated flux densities. The fraction of the area under this distribution
for $|r|\geq r_{\rm data}$, where $r_{\rm data}$ is the $r-$value for the
observed flux densities, is the probability to have obtained an apparent
correlation at least as strong as a the one seen in the data from a sample with
intrinsically uncorrelated gamma-ray/radio emission. This quantifies the
statistical significance of the observed correlation.

Our results for all the correlations discussed in the present paper are shown
in Tables \ref{t.sigAll} (full 1LAC sample), \ref{t.sigOVROc} (OVRO sample,
using concurrent radio data), \ref{t.sigOVROnc} (OVRO sample, using
non-concurrent radio data), and \ref{t.sig4bands} (sample of sources detected
in at least 4 bands); for every case examined, we give the number of sources in
the studied subset, the number of redshift bins used in the analysis, the
Spearman correlation coefficient $\rho$, the value of the Pearson correlation
coefficient $r$ of the dataset, and the statistical significance of the
apparent correlation, which we define as the fraction of simulated datasets
with the same number of points, same common-distance, luminosity-range, and
flux-range effects as the actual dataset {\em but no intrinsic correlation}
which had an absolute value of $r$ at least as big as the actual dataset. The
number of points in each dataset studied generally differs from the number of
points in the corresponding dataset of \S \ref{s.1lac}, \ref{s.ovro} and
\ref{s.4bands}, because for the surrogate data studies we only use sources for
which the redshifts are known. In the scatter plots of
Figs.~\ref{f.1lac-sum}--\ref{f.1lac-sed},
\ref{f.ovro15-sum}--\ref{f.ovro15-sed}, and
\ref{f.4bands-sum}--\ref{f.4bands-sed}, these sources are plotted with black
points (while magenta is used for sources with unknown $z$). For the same
reason, the Spearman correlation coefficient for the sets submitted to the
surrogate data analysis is also different but consistent with the value shown
in Tables \ref{t.1lac}-\ref{t.4bands}.

In the case of large samples and relatively high correlation coefficients, the
apparent correlations between radio and gamma-ray flux are found to be also
{\it intrinsically} very significant: for example, the probability of the
correlation between $E>100$~MeV gamma-ray flux and the 8\,GHz archival data
arising due to common-distance effects or the limited flux and luminosity
ranges examined is smaller than $10^{-7}$. However, for smaller subsets and
weaker correlations (lower values of the correlation coefficients) the
significance of the correlation cannot be established with such high confidence
or not at all. A striking example is that of the correlation between 8\,GHz
archival flux densities and 10-100 GeV fluxes four sources that were detected
at least in four bands: the simulated datasets are more strongly correlated
than the actual dataset more than 30\% of the time.

\begin{figure}
\plotone{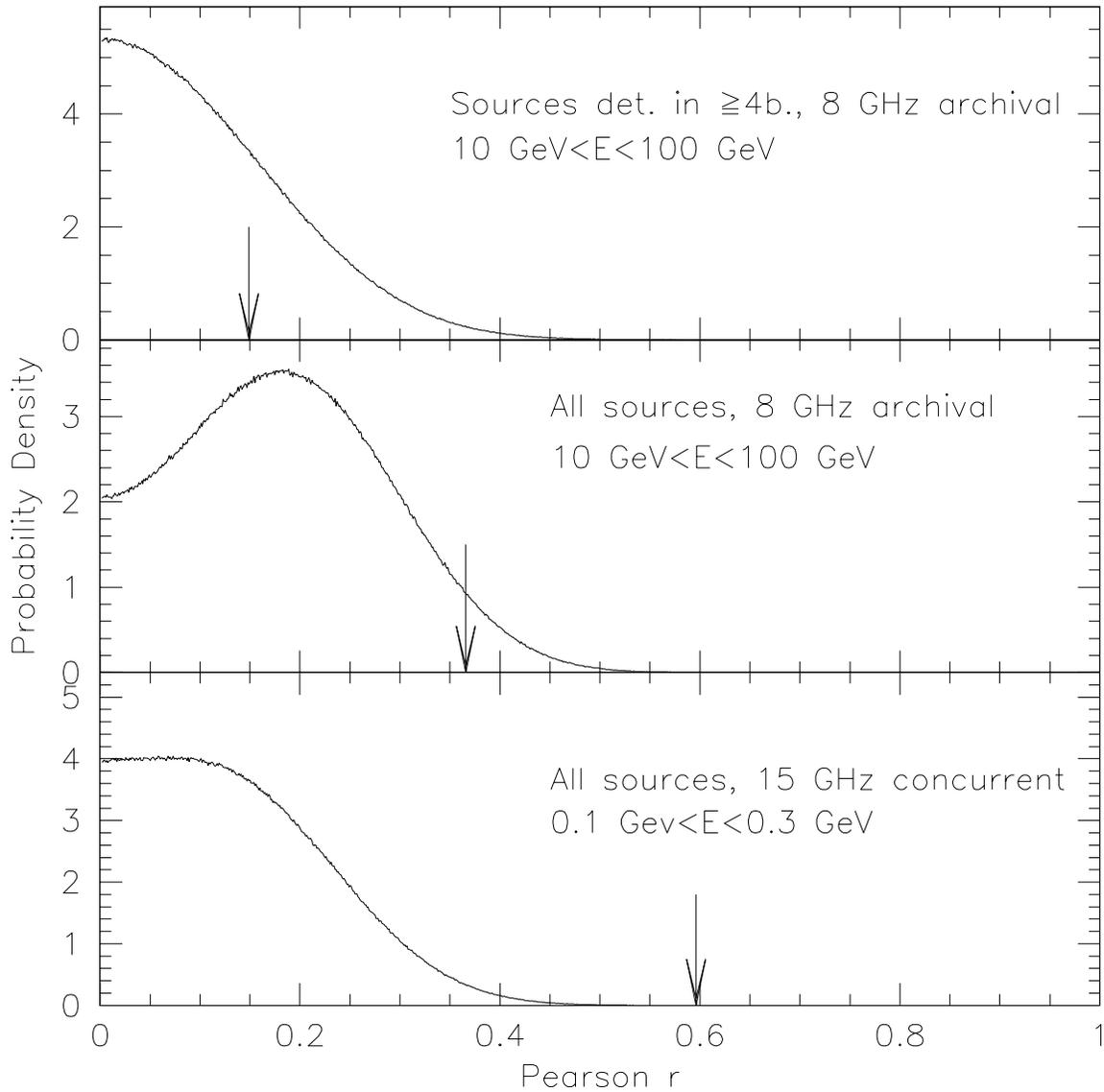}
\caption{Probability density distributions of the absolute value of the Pearson
  product-moment $r$ for three simulated datasets, with low (top panel), medium
  (middle), and high (bottom) correlation significance.  \label{f.significances}}
\end{figure}

In Fig.~\ref{f.significances} we show, for three example cases, the
distribution of the absolute value of the Pearson product-moment $r$ for
simulated datasets. The value of $r$ for the actual dataset in each case is
indicated with an arrow. The three cases are selected so that they represent
examples of low (top panel), medium (middle panel) and high (bottom panel)
correlation significances. In some cases (as in the middle panel of these
examples) the distribution of $r$ for simulated, intrinsically uncorrelated
datasets peaks at a finite positive value {\em even in flux space}. This
generally indicates a clustering of intrinsic luminosities around a specific
value; these luminosities then are more frequently selected, even by chance,
and when a common redshift is applied to them they result in a positive
correlation coefficient, which then becomes more common than the zero value
\citep[for a more detailed discussion of this effect, see also][]{Methodology}.

The interpretation of the quoted significances requires some care, owing to the
large number of subsamples (a total of 144) examined using the data shuffling
technique. The quoted significances are a useful tool in comparing the
significance of any apparent correlation between different subsamples, however
the correlation significance in any single subsample is severely mitigated
because of the issue of trials. E.g., any effect that occurs 5\% of the time by
chance would have a probability of 99.9\% to occur at least once in 144
independent trials; this probability becomes 76.3\% for effects occurring by
chance 1\% of the time, and 1\% for effects occurring by chance only $10^{-4}$
of the time. Although the subsamples considered are, in fact, not independent,
the numbers above may serve as illustration that interpreting the here quoted
significances at "face-value" can be misleading. However, for the largest
samples considered here, we have found correlations with very high significance
($<10^{-7}$), which remain very confident even in the face of the large number
of trials.

We further remind that the method of surrogate data discussed above is
applicable only to samples for which the redshifts of the sources are
known. For this reason, when calculating significances for the apparent
correlations of various subsamples, we discard sources for which the redshift
is not known (most of such sources are BL Lacs). However, the omission of
sources without known redshifts can affect the evaluated significance in two
ways: by altering the redshift distribution of the sample, and by reducing the
number of sources. As a quantitative example of these effects, we have tested
how our calculated significances change if we include the sources without known
redshifts, and we assign redshifts to them in the following two ways: (a)
assume that the missing redshifts have the same distribution as the known
redshifts; in this case, we randomly select a redshift from sources of the same
type (in most cases, BL Lacs) for the sources without redshifts; (b) assume
that the missing redshifts have systematically higher values than the known
redshifts; in particular, we assume that the distribution of the missing
redshifts is that of known redshifts (for sources of the same type) translated
to higher redshifts by $\Delta z=0.5$. We have tested these two cases for the
sample of HSP blazars using 8 GHz radio flux densities and gamma-ray flux in
the 10-100 GeV band.

\begin{figure}
\plotone{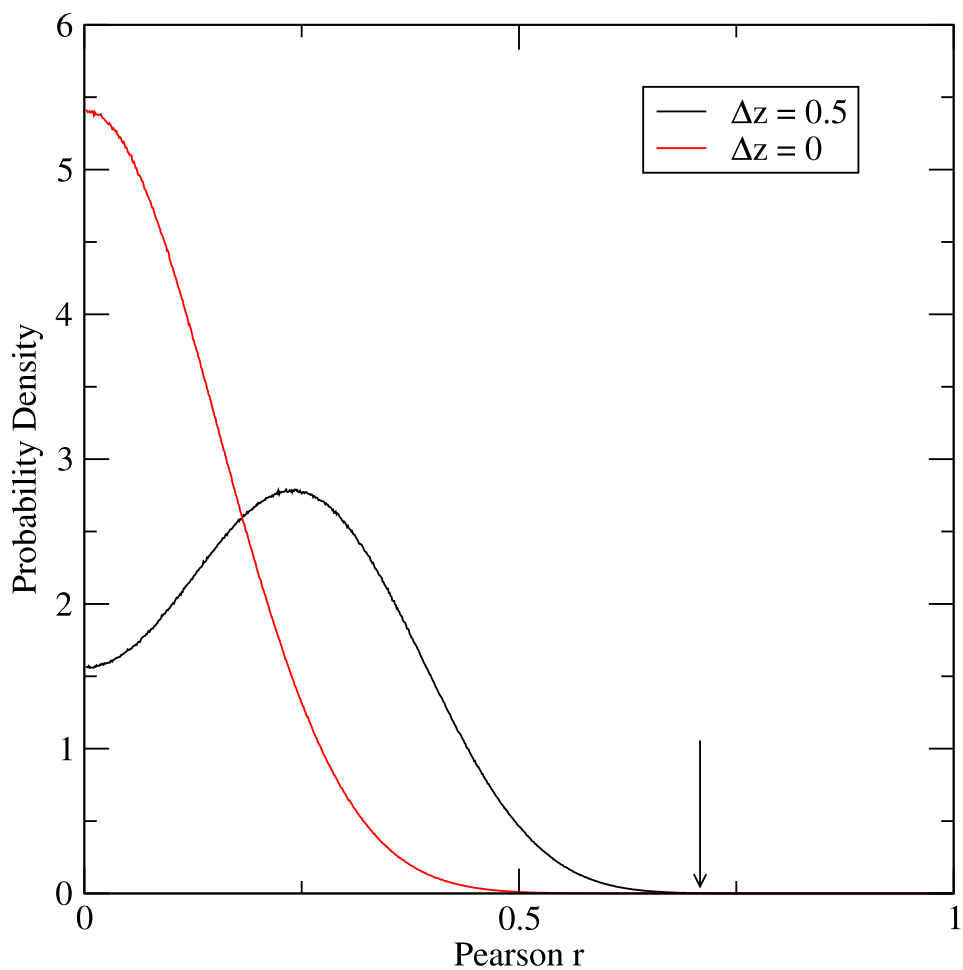}
\caption{Probability density distribution of the absolute value of the Pearson product-moment $r$ for HSP blazars using 8 GHz archival data and 10--100 GeV gamma-ray flux density, assuming that the sources without redshift follow the same redshift distribution of the ones with known $z$ (red solid line) or with a mean shift of $\Delta z = 0.5$ (black solid line).  \label{f.zstudy}}
\end{figure}

The results for these test cases are shown in Fig.~\ref{f.zstudy}. As an effect
of the increased number of sources available for the test (47 instead of 21),
the significance increases from $2.0\times 10^{-2}$ to $4.2\times10^{-5}$ (if
the sources without redshift follow the distribution generated with $\Delta z =
0.5$), or even to $3.0 \times 10^{-8}$ (if they are distributed in the same way
as the sources with a measured redshift). Note however that the results are
substantially different depending on our choice of how to assign simulated
redshifts, and for this reason we have not implemented this technique more
extensively in our samples.

\section{Discussion \label{s.discussion}}

The sources in the 1LAC sample have an overall positive correlation between
radio flux density and gamma-ray photon flux, with a very high statistical
significance as supported by the dedicated statistical analysis presented in
Sect.~\ref{s.mc}. Moreover, the vast majority of the statistical tests run on
the distribution of the gamma-ray and radio flux densities for the various
source type/energy range combinations has also revealed some correlation
(Sect.~\ref{s.results}), with moderate-to-high statistical
significance. Overall, this confirms the existence of a relationship between
the emission in these two distant parts of the electromagnetic spectrum. This
finding is consistent with other studies on the subject
\citep{Kovalev2009,Ghirlanda2010,Mahony2010}. Most imporantly, it has now been
demonstrated to be robust against common-distance effects, and the effect of a
limited flux and luminosity range.

In addition, the sensitivity of the LAT over three decades in energy range
allows us to characterize a huge number of extragalactic gamma-ray sources
across the gamma-ray band and to clarify some details of the relationship. The
quality of the radio data provided by the archives as well as from concurrent
monitoring are also crucial for a better understanding of the general picture.

For instance, BL Lacs are under represented in analyses performed starting from
samples with moderate or high radio flux density limits, such as the AT20GHZ
\citep{Ghirlanda2010,Mahony2010} and the MOJAVE \citep{Kovalev2009}, whereas
they actually constitute more than half of the 1LAC. Thanks to the archival
interferometric data obtained for the full sample, we have studied the
radio/gamma-ray connection within the two blazar sub-populations separately
with a large number of sources. Indeed, even when considered independently,
1LAC BL Lacs display a correlation between their radio and gamma-ray flux
densities that is highly significant; the chance probability is, e.g.,
$<10^{-7}$ when considering the full energy band, and $1.9\times 10^{-6}$ in
the 1--3 GeV energy band (see Table \ref{t.sigAll}). As the surrogate data
method can only be applied to sources with a known distance, it would be
desirable to have more redshifts available for BL Lacs in order to improve the
significance of this correlation also for other sub-bands. However, even with
the lack of more redshift measurements, the conclusions implied from
Fig.~\ref{f.zstudy} make one expect that such significance is no less than that
of FSRQs, or even higher given the larger value of both Spearman's $\rho$ and
Pearson $r$ for BL Lacs.

The finding of a high apparent correlation strength for BL Lacs is not only
present in the full 1LAC, but is also present -- and actually with higher
values of $\rho$ -- when one considers the results obtained for the sample of
sources detected in at least four bands (\S\ref{s.4bands}). This sub-sample
probably provides the most robust results, for two reasons. First, the $\rho$
values are obtained by considering largely overlapping samples in each energy
bin; second, since these are moderately bright gamma-ray sources (they would
not be significant in 4/5 energy bands otherwise), the fluxes are better
constrained and we are not too close to the detection limit. Still, we caution
that the 4-band sample may not be a fully representative sample of the whole
gamma-ray sky, as it is about 1/3 of the 1LAC.

As far as the OVRO sample is concerned, it seems to have yielded somewhat
different details of the overall picture with respect to the full 1LAC. This is
not entirely surprising, as the two samples represent different populations:
the OVRO sample is generally brighter compared to the whole 1LAC set, and FSRQs
are more strongly represented. In any case, the availability of the large,
long-term, high-cadence monitored OVRO sample is of great importance in the
assesment of the role of variability on the radio-gamma connection. The OVRO
data clearly reveal, for the first time, that concurrent radio fluxes are more
strongly correlated with gamma-ray fluxes than archival data, even at the same
frequency. For example, the significance of the correlation between radio and
gamma-ray broad band fluxes for all sources increases from $1.9\times 10^{-6}$
to $9.0\times 10^{-8}$ when going from non-simultaneous to concurrent
data. Increased signficance is found also for most of the various combinations
of source type and energy band. This was a long-expected result, which has
finally been demonstrated. Interestingly, the peak radio flux density during
the time of collection of gamma-ray data shows a weaker correlation than the
one obtained using the mean values; it is actually even weaker than that of
non-concurrent radio data. The fact that the the strongest correlation is
obtained for the time-integral of the flux density in the two bands shows that
the best correlation is between the overall energy dissipated in the two
regimes.

A further advantage of such a large dataset, which distinguishes our results
from past work, is that it provides sufficient number of sources for good
statistical analysis, even when we divide the sample in finer sub-groups, for
example on the basis of the spectral properties in the synchrotron component of
the SED. The possibility of sub-grouping is interesting, particularly when we
compare the results obtained dividing by optical type and position of the peak
of the low-energy component and/or considering each of the individual LAT
energy bands. However, even if we are for the first time in the position of
attempting such studies, we have to keep in mind that the statistical
significance becomes inevitably lower when the samples are less populated, so
the following discussion is certainly somewhat speculative.

First, the BL Lacs seem to follow a pattern of lower correlation coefficients
when gamma-rays of increasing energy are considered -- with flux densities that
become apparently uncorrelated (or even anti-correlated) in the highest energy
band; this is particularly prominent in the OVRO sample. However, when the
spectral types are considered separately, a pattern emerges with HSP always
being the class with the strongest correlation in the 1--3, 3--10, and 10-100
GeV energy bands. ISP BL Lacs, on the other hand, show much weaker or absent
correlation at high energy, which affects the total population of BL Lacs when
considered as a whole. This effect becomes most prominent when the ISP/HSP
population ratio is higher.

Second, LSP blazars are more difficult to characterize, since they are a mixed
population of both BL Lacs and FSRQs. We note that FSRQ and LSP however do not
always follow the same trend. It is thus likely that the radio and gamma-ray
emission in FSRQ-LSP and the BL Lac-LSP are not produced by the same kind of
process. This may or may not be related to the other well known differences in
the optical spectrum and in the accretion regime for these two populations.

The interpretation of the dependence of the correlation strength on the source
type is therefore in general not straightforward. The fact that HSP sources
show the strongest correlation could be related to the fact that these sources
do not generally possess large amounts of extended emission, and even on parsec
scales their jets are rather weak. So the interferometric flux density is
probably more representative of the properties of the region where the
gamma-rays are produced. Moreover, in HSP sources the high energy component of
the SED extends to the TeV band, so that the particles involved in the radio
and GeV emission could be low energy electrons, unlike the LSP case where the
GeV emission requires high energy particles.

The physical reason why even within a single population the correlation
coefficient is quite strongly dependant on the considered energy band is even
more difficult to interpret. For example, ISP blazars have a peak of
correlation in the low energy band, while their radio and gamma-ray flux
densities become essentially uncorrelated at the highest energies:\ sources of
any radio flux density seem to produce more or less the same amount of
gamma-ray photons. In other words, the radio-bright ISP blazars would have much
softer gamma-ray spectra than the radio weak, which would be somewhat
consistent with the picture in the blazar sequence \citet{Fossati1998};
however, the same trend is not observed in other blazar classes.

Finally, one should be cautioned against correlations that are driven from a
minority of very high (or very low) data points. For instance,
\citet{Linford2011} find a flux-flux correlation for 50 gamma-ray FSRQs in the
VLBA Imaging and Polarimetry Survey \citep[VIPS,][]{Helmboldt2007}, but they
also discover that the correlation disappears when the 10\% brightest sources
in the radio are discarded. In our 247 FSRQs sample, the effect is not quite as
dramatic, with only a modest decrease of the correlation coefficient. As BL
Lacs are entirely unaffected, however, this could still be an interesting clue
about additional differences between the two classes.

\subsection{Luminosity Distributions}

Throughout this paper, we discuss the strength and the significance of
radio/gamma correlations in terms of the flux density in each band. We do not
discuss luminosity correlations, as the limited dynamical range in fluxes,
combined with the aggregation of sources close to the flux limit and the
square-distance effect always induce a strong apparent correlation in
luminosity space, whereas plots in the flux density plane give a better visual
impression of the scatter involved. Moreover, it is possible to show
analytically that the information in the flux/flux and luminosity/luminosity
correlations is essentially degenerate.

However, from the physical point of view it is interesting to examine also the
ranges and distributions of luminosities in the gamma-ray band and in the radio
band, which are shown in \citet{1LAC}.  The luminosity ranges probed by our
sample extend over approximately two orders of magnitude in each band for BL
Lacs and FSRQs, while they are more extended for other AGN.  Although there is
significant overlap between the luminosities of different source types, FSRQs
are generally more luminous than BL Lacs, which are in turn more luminous than
the remaining AGNs (which includes radio galaxies and blazars of uncertain
type); this result holds both for the radio and the gamma-ray bands. Therefore,
we suggest that there is a lot of discovery space for sources of low gamma-ray
luminosities, as the luminosity range of unclassified AGN and radio galaxies
extends more than two orders of magnitude fainter than BL Lacs and FSRQ in the
gamma-ray band, but it is so far much less populated than the higher luminosity
domain. It is possible that some of the fainter unassociated gamma-ray sources
at high latitudes are AGN in this luminosity range.

\section{Conclusions \label{s.conclusions}}

We have searched for a possible intrinsic correlation between gamma-ray and
radio fluxes. We have found that such a correlation does exist, and it is
statistically significant for the largest sample we have studied that includes
all source types: the probability that it arises by chance (e.g., through
common-distance effects, accentuated by the limited dynamical range of fluxes
of the sample) is smaller than $10^{-7}$. The significance is also very high
when FSRQ and BL Lacs are considered independently. However, the distribution
of sources along the correlation has appreciable scatter (which can be
typically an order of magnitude). Therefore, we strongly caution that any use
of this intrinsic connection between radio and gamma-ray emission in
statistical descriptions of the gamma-ray population, such as to obtain
gamma-ray luminosity functions from radio luminosity functions, should be done
with care and always accounting for the scatter involved. When comparing
archival with concurrent data we find that the moderate significance of a
correlation derived from the archival radio -- gamma-ray sample increases
appreciably when concurrent data are used.

The statistical significance of a correlation does not have a simple dependence
on the {\em apparent correlation strength}. Various other factors play a role
in an assessment of the significance of an apparent correlation, besides the
tightness of the observed correlation itself. These include the following: (1)
errors in the observed fluxes, (2) any biases, e.g., the presence of
common-distance effects and flux limits, and (3) the number of sources in the
sample. While underlying statistical errors (1) are inherent, in the present
work we have explicitly accounted for biases (2). The sample size (3) affects
the importance of "cosmic variance": a small sample with a significant
correlation but appreciable statistical errors or scatter might happen to
appear uncorrelated, and, conversely, any single incarnation of a small,
uncorrelated sample might appear correlated by chance.  Considering all of
these factors, we have established high significance for some correlations but
the same is not possible for very small subsamples.

We have studied the radio/gamma correlation for different subsets of blazars,
and we have found that the {\em apparent strength} of the correlation depends
on the type of blazar; in particular, BL Lacs have been shown to possess a high
apparent correlation strength using the largest sample ever considered. The
{\em apparent strength} of the correlation depends also on the epoch of
observation, in that concurrent radio and gamma-ray measurements correlate
better than data obtained at different epochs in the two bands.

Finally, the specific gamma-ray energy band over which the gamma-ray flux is
calculated seems also to affect the strength of the correlation, with HSP
blazars generally displaying a stronger correlation. The highest apparent
correlation strength appears at higher gamma-ray energy for HSPs than for LSPs
and ISPs. Both the specific energy bands and source types considered impact
also the significance of the correlation. These results have been obtained
thanks to the large number of AGNs detected by the {\em Fermi}-LAT, although in
some of the considered correlations the number of sources is small. Therefore,
a further increase in the number of objects in each sub-group, as that expected
for the 2LAC, is needed to improve the significance of individual results.

\acknowledgments The \textit{Fermi} LAT Collaboration acknowledges generous
ongoing support from a number of agencies and institutes that have supported
both the development and the operation of the LAT as well as scientific data
analysis. These include the National Aeronautics and Space Administration and
the Department of Energy in the United States, the Commissariat \`a l'Energie
Atomique and the Centre National de la Recherche Scientifique / Institut
National de Physique Nucl\'eaire et de Physique des Particules in France, the
Agenzia Spaziale Italiana and the Istituto Nazionale di Fisica Nucleare in
Italy, the Ministry of Education, Culture, Sports, Science and Technology
(MEXT), High Energy Accelerator Research Organization (KEK) and Japan Aerospace
Exploration Agency (JAXA) in Japan, and the K.~A.~Wallenberg Foundation, the
Swedish Research Council and the Swedish National Space Board in Sweden.

Additional support for science analysis during the operations phase is
gratefully acknowledged from the Istituto Nazionale di Astrofisica in Italy and
the Centre National d'\'Etudes Spatiales in France.

{\it Facilities:} \facility{{\it Fermi} LAT}, \facility{OVRO:40m ()}, \facility{VLA}, \facility{ATCA}

\bibliography{radiogamma}

\clearpage

\begin{deluxetable}{llrrccrrcrrr}
\rotate
\tablecolumns{11} 
\tabletypesize{\scriptsize}
\tablecaption{\label{t.radiodata}Source list and radio data.}  
\tablewidth{0pt}
\tablehead{
\colhead{1FGL name} &
\colhead{Other name} &
\colhead{RA} &
\colhead{Dec} &
\colhead{Opt.\ type} &
\colhead{SED type} &
\colhead{$z$} &
\colhead{$S_8$} &
\colhead{Ref.} &
\colhead{$S_\mathrm{15,\, 1yr\, mean}$} &
\colhead{$S_\mathrm{15,\, 1yr\, peak}$} &
\colhead{$S_\mathrm{15,\, non\, c.}$} \\
\colhead{} &
\colhead{} &
\colhead{($^\circ$)} &
\colhead{($^\circ$)} &
\colhead{} &
\colhead{} &
\colhead{} &
\colhead{(mJy)} &
\colhead{} &
\colhead{(mJy)} &
\colhead{(mJy)} &
\colhead{(mJy)} 
}
\startdata
1FGL\,J0000.9$-$0745 & CRATES\,J0001$-$0746 &       0.32512 &  $-$7.77417 &            BLL   &      LSP   &  \nodata  &     116.2  &   Cr   &        \nodata  &    \nodata  &    \nodata\\
1FGL\,J0004.7$-$4737 & PKS\,0002$-$478      &       1.14867 & $-$47.60517 &           FSRQ   &      LSP   &     0.88  &     780.4  &   Cr   &        \nodata  &    \nodata  &    \nodata\\
1FGL\,J0005.7+3815   & B2\,0003+38A         &       1.48825 &  38.33755 &           FSRQ   &      LSP   &    0.229  &      1078  &   Cr   &       541.9    &   703.3    &   541.2 \\
1FGL\,J0008.9+0635   & CRATES\,J0009+0628   &       2.26638 &   6.47256 &            BLL   &      LSP   &  \nodata  &     196.5  &   Cr   &        \nodata  &    \nodata  &    \nodata\\
1FGL\,J0011.1+0050   & CGRaBS\,J0011+0057   &       2.87667 &   0.96439 &           FSRQ   &      LSP   &    1.492  &     278.7  &   Cr   &       249.3    &   364.8    &   205.6 \\
1FGL\,J0013.1$-$3952 & PKS\,0010$-$401      &       3.24962 & $-$39.90717 &            BLL   &      \nodata   &  \nodata  &      1554  &   Cr   &        \nodata  &    \nodata  &    \nodata\\
1FGL\,J0013.7$-$5022 & BZB\,J0014$-$5022    &       3.54675 & $-$50.37575 &            BLL   &      HSP   &  \nodata  &      13.4  &   N      &        \nodata  &    \nodata  &    \nodata\\
1FGL\,J0017.4$-$0510 & CGRaBS\,J0017$-$0512 &       4.39925 &  $-$5.21158 &           FSRQ   &      LSP   &    0.227  &     225.2  &   Cr   &       297.8    &   497.3    &   173.1 \\
\enddata
\tablecomments{The optical data reported in this table are taken directly from the 1LAC \citep{1LAC} and have been used in the data shuffling. We note that the measure of $z$ for a few sources has improved \citep[e.g.\ PG\,1553+113,][]{Danforth2010} and that for local sources (e.g.\ Cen\,A) the redshift information does not necessarily provide the most accurate estimate of the actual distance.}
\tablerefs{Cr:\ CRATES \citep{crates}; Cl:\ CLASS \citep{class1,class2}; F:\ FRBA (Finding and Rejecting Blazar Associations for Fermi-LAT gamma-ray sources, from 8.4 GHz VLA program AH996); N:\ data extrapolated from NED/NVSS/Bzcat; A95:\ \citet{Akujor95}; L97:\ \citet{Laurent97}; N01: \citet[][$S_8=2.92$\,Jy used in the calculation]{Nagar01}; T03: \citet[][$S_8=4.99$\,Jy and $S_8=2.95$\,Jy used in the calculation for 1FGL\,J0522.8-3632 and 1FGL\,J1325.6-4300, respectively]{Tingay03}; U84:\ \citet{Unger84}.}
\end{deluxetable}

\begin{deluxetable}{lcccccccccccc}
\tablecolumns{13} 
\tabletypesize{\footnotesize}
\tablecaption{\label{t.1lac}Spearman rank correlation coefficient $\rho$ for
  the 1LAC clean sample, divided by source type. For each energy band ($E$
  measured in GeV), the number of significant sources $n$ and the correlation
  coefficient $\rho$ are given.}  
\tablewidth{0pt}
\tablehead{
\colhead{Source class} & 
\multicolumn{2}{c}{$E>0.1$}  & 
\multicolumn{2}{c}{$0.1<E<0.3$} & 
\multicolumn{2}{c}{$0.3<E<1$} & 
\multicolumn{2}{c}{$1<E<3$} & 
\multicolumn{2}{c}{$3<E<10$} & 
\multicolumn{2}{c}{$10<E<100$} \\
\colhead{} &
\colhead{$n$} & \colhead{$\rho$} &
\colhead{$n$} & \colhead{$\rho$} &
\colhead{$n$} & \colhead{$\rho$} &
\colhead{$n$} & \colhead{$\rho$} &
\colhead{$n$} & \colhead{$\rho$} &
\colhead{$n$} & \colhead{$\rho$}
}
\startdata
All sources & 599 & 0.43 & 193 & 0.41 & 384 & 0.51 & 493 & 0.43 & 373 & 0.31 & 127 & 0.30 \\ 
\hline
FSRQ        & 248 & 0.39 & 128 & 0.32 & 214 & 0.41 & 220 & 0.35 & 140 & 0.29 & 22 & 0.42 \\ 
BL Lacs     & 275 & 0.46 & 47  & 0.48 & 135 & 0.54 & 214 & 0.54 & 197 & 0.35 & 99 & 0.31 \\ 
\hline
LSP         & 242 & 0.40 & 133 & 0.37 & 200 & 0.42 & 219 & 0.38 & 158 & 0.39 & 35 & 0.35 \\ 
ISP         & 60  & 0.33 & 18  & 0.58 &  41 & 0.40 & 53  & 0.44 & 42  & 0.29 & 15 & 0.30 \\ 
HSP         & 129 & 0.55 & 12  & 0.16 &  47 & 0.62 & 87  & 0.64 & 88  & 0.57 & 64 & 0.55 \\ 
\enddata
\end{deluxetable}

\begin{deluxetable}{lcccccccccccc}
\tablecolumns{13} \tabletypesize{\footnotesize}
\tablecaption{\label{t.ovro}Spearman rank correlation coefficient $\rho$ for
  the subset of the 1LAC clean sample monitored by OVRO, divided by source
  type. For each energy band ($E$ measured in GeV), the number of significant
  sources $n$ and the correlation coefficient $\rho$ are given.}

\tablewidth{0pt}
\tablehead{
\colhead{Source class} & 
\multicolumn{2}{c}{$E>0.1$}  & 
\multicolumn{2}{c}{$0.1<E<0.3$} & 
\multicolumn{2}{c}{$0.3<E<1$} & 
\multicolumn{2}{c}{$1<E<3$} & 
\multicolumn{2}{c}{$3<E<10$} & 
\multicolumn{2}{c}{$10<E<100$} \\[3pt]
\colhead{} &
\colhead{$n$} & \colhead{$\rho$} &
\colhead{$n$} & \colhead{$\rho$} &
\colhead{$n$} & \colhead{$\rho$} &
\colhead{$n$} & \colhead{$\rho$} &
\colhead{$n$} & \colhead{$\rho$} &
\colhead{$n$} & \colhead{$\rho$}
}
\startdata
\multicolumn{13}{c}{8 GHz non simultaneous radio data} \\
All sources & 199 & 0.27 & 97 & 0.40 & 163 & 0.41 & 179 & 0.25 & 134 & 0.15 & 40 & $-$0.04 \\ 
\hline
FSRQ        & 120 & 0.38 & 67 & 0.36 & 104 & 0.41 & 109 & 0.36 & 73 & 0.27 & 13 & 0.38 \\ 
BL Lacs     & 69  & 0.08 & 24 & 0.30 & 51  & 0.28 & 61  & 0.14 & 57 & 0.03 & 25 & $-$0.19 \\ 
\hline
LSP         & 131 & 0.30 & 77 & 0.42 & 113 & 0.41 & 120 & 0.29 & 89 & 0.28 & 24 & 0.16 \\ 
ISP         & 17  & 0.07 & 6  & 0.09 & 16 & 0.37 & 16 & 0.09 & 13 & $-$0.19 & 4 & 0.00 \\ 
HSP         & 10  & 0.73 & 6  &$-$0.02 & 10 & 0.64 & 10 & 0.79 & 10 & 0.67 & 9 & 0.67 \\ 
\hline
\hline
\multicolumn{13}{c}{15 GHz concurrent radio data, 1-yr mean values} \\
\enddata
\end{deluxetable}

\begin{deluxetable}{lcccccccccccc}
\tablecolumns{13} 
\tabletypesize{\footnotesize}
\tablecaption{\label{t.4bands}Spearman rank correlation coefficient $\rho$ for
  1LAC sources with a detection in at least 4 energy bands, divided by source type. For each energy band ($E$
  measured in GeV), the number of significant sources $n$ and the correlation
  coefficient $\rho$ are given.}  

\tablewidth{0pt}
\tablehead{
\colhead{Source class} & 
\multicolumn{2}{c}{$E>0.1$}  & 
\multicolumn{2}{c}{$0.1<E<0.3$} & 
\multicolumn{2}{c}{$0.3<E<1$} & 
\multicolumn{2}{c}{$1<E<3$} & 
\multicolumn{2}{c}{$3<E<10$} & 
\multicolumn{2}{c}{$10<E<100$} \\
\colhead{} &
\colhead{$n$} & \colhead{$\rho$} &
\colhead{$n$} & \colhead{$\rho$} &
\colhead{$n$} & \colhead{$\rho$} &
\colhead{$n$} & \colhead{$\rho$} &
\colhead{$n$} & \colhead{$\rho$} &
\colhead{$n$} & \colhead{$\rho$}
}
\startdata
All sources & 192 & 0.45 & 149 & 0.45 & 192 & 0.60 & 192 & 0.44 & 192 & 0.15 & 94 & 0.21 \\ 
\hline
FSRQ        & 94  & 0.30 & 93  & 0.34 & 94  & 0.30 & 94 & 0.24 & 94 & 0.15 & 22 & 0.42 \\ 
BL Lacs     & 84  & 0.55 & 44  & 0.48 & 84  & 0.72 & 84 & 0.62 & 84 & 0.41 & 67 & 0.23 \\ 
\hline
LSP         & 107 & 0.40 & 102 & 0.42 & 107 & 0.39 & 107 & 0.33 & 107 & 0.24 & 33 & 0.29 \\ 
ISP         & 20  & 0.46 & 15  & 0.52 & 20  & 0.63 & 20  & 0.26 & 20  & 0.16 & 12 & 0.02 \\ 
HSP         & 41  & 0.71 & 12  & 0.16 & 41  & 0.65 & 41  & 0.74 & 41  & 0.64 & 40 & 0.64 \\ 
\enddata
\end{deluxetable}

\begin{deluxetable}{llccccc}
\tablecolumns{7} 
\tabletypesize{\scriptsize}
\tablecaption{\label{t.sigAll} Significances of intrinsic correlations after common-distance effects are taken into account, for sources with known redshift, using 8\,GHz archival data, as a function of source type and $\gamma-$ray energy band. }  
\tablewidth{0pt}
\tablehead{
\colhead{Source type} & \colhead{Energy Band} & 
\colhead{Number of sources} & \colhead{Number of z-bins} & \colhead{Spearman $\rho$} &\colhead{Pearson $r$} &
\colhead{significance}
}
\startdata
All sources &$E>0.1$    & 390 & 10 & 0.46 & 0.47 & $<10^{-7}$ \\
            &$0.1<E<0.3$& 164 & 10 & 0.36 & 0.42 & $3.4\times 10^{-5}$ \\
            &$0.3<E<1$  & 281 & 10 & 0.45 & 0.48 & $<10^{-7}$ \\
            &$1<E<3$	& 329 &	10 & 0.42 & 0.44 & $<10^{-7}$ \\
            &$3<E<10$   & 235 & 10 & 0.34 & 0.36 & $8.1\times 10^{-6}$ \\
            &$10<E<100$ & 72  & 5  & 0.34 & 0.37 & 0.046 \\
\hline
FSRQ        &$E>0.1$    & 248 & 10 & 0.39 & 0.42 & $<10^{-7}$ \\ 
            &$0.1<E<0.3$& 128 &	10 & 0.32 & 0.40 & $1.13\times 10^{-4}$ \\
\enddata
\end{deluxetable}

\begin{deluxetable}{llccccc}
\tablecolumns{7} 
\tabletypesize{\scriptsize}
\tablecaption{\label{t.sigOVROc} Significances of intrinsic correlations after common-distance effects are taken into account, for sources in the OVRO sample with known redshift, using 15\,GHz concurrent data, as a function of source type and $\gamma-$ray energy band. }  
\tablewidth{0pt}
\tablehead{
\colhead{Source type} & \colhead{Energy Band} & 
\colhead{Number of sources} & \colhead{Number of z-bins} & \colhead{Spearman $\rho$} &\colhead{Pearson $r$} &
\colhead{significance}
}
\startdata
All sources &$E>0.1$    & 160 & 10 & 0.39 & 0.46 & $9.0\times 10^{-8}$ \\
            &$0.1<E<0.3$& 87  & 5  & 0.55 & 0.60 & $1.3\times 10^{-6}$ \\   
            &$0.3<E<1$  & 136 & 10 & 0.50 & 0.57 & $3.0\times 10^{-8}$ \\   
            &$1<E<3$	& 146 & 10 & 0.40 & 0.47 & $2.7\times 10^{-6}$ \\	  
            &$3<E<10$   & 104 & 10 & 0.22 & 0.28 & 0.021 \\
            &$10<E<100$ & 29  & 3  & 0.06 & 0.03 & $0.89$ \\                
\hline                                                                      
FSRQ        &$E>0.1$    & 120 & 10 & 0.48 & 0.53 & $1.4\times 10^{-7}$ \\
            &$0.1<E<0.3$& 67  & 6  & 0.56 & 0.61 & $9.0\times 10^{-8}$ \\
\enddata
\end{deluxetable}

\begin{deluxetable}{llccccc}
\tablecolumns{7} 
\tabletypesize{\scriptsize}
\tablecaption{\label{t.sigOVROnc} Significances of intrinsic correlations after common-distance effects are taken into account, for sources in the OVRO sample with known redshift, using 15\,GHz non concurrent data, as a function of source type and $\gamma-$ray energy band. }  
\tablewidth{0pt}
\tablehead{
\colhead{Source type} & \colhead{Energy Band} & 
\colhead{Number of sources} & \colhead{Number of z-bins} & \colhead{Spearman $\rho$} &\colhead{Pearson $r$} &
\colhead{significance}
}
\startdata
All sources &$E>0.1$    & 160 & 10 &      0.36 &      0.42 & $1.9\times 10^{-6}$\\
            &$0.1<E<0.3$&  87 & 5  &      0.47 &      0.51 & $7.2\times 10^{-5}$\\
            &$0.3<E<1$  & 136 & 10 &      0.45 &      0.50 & $2.0\times 10^{-7}$\\    
            &$1<E<3$	& 146 & 10 &      0.36 &      0.42 & $4.1\times 10^{-5}$\\    
            &$3<E<10$   & 104 & 10 &      0.21 &      0.25 & 0.045 \\
            &$10<E<100$ &  29 & 2 &     $-$0.06 &     $-$0.05 &	 0.78\\                  
\hline                                                                                     
FSRQ        &$E>0.1$    & 120 & 10 &      0.45 &      0.50 & $1.0\times 10^{-6}$\\
            &$0.1<E<0.3$&  67 & 6 &      0.49 &      0.54  & $2.5\times 10^{-6}$\\
\enddata
\end{deluxetable}

\begin{deluxetable}{llccccc}
\tablecolumns{7} 
\tabletypesize{\scriptsize}
\tablecaption{\label{t.sig4bands} Significances of intrinsic correlations after common-distance effects are taken into account, for sources detected in at least 4 bands, with known redshift, using 8\,GHz archival data, as a function of source type and $\gamma-$ray energy band.}  
\tablewidth{0pt}
\tablehead{
\colhead{Source type} & \colhead{Energy Band} & 
\colhead{Number of sources} & \colhead{Number of z-bins} & \colhead{Spearman $\rho$} &\colhead{Pearson $r$} &
\colhead{significance}
}
\startdata
All sources &$E>0.1$    & 138	&	10 &  0.33 &	  0.38 & $1.2\times 10^{-3}$ \\
            &$0.1<E<0.3$& 125   &       10 &  0.39 &      0.44 & $1.4\times 10^{-4}$ \\
            &$0.3<E<1$  & 138	& 	10 &  0.42 &      0.50 & $1.5\times 10^{-5}$ \\
            &$1<E<3$	& 138	&	10 &  0.28 &      0.36 & $3.3\times 10^{-3}$ \\
            &$3<E<10$   & 138	&	10 &  0.12 &      0.19 &  0.21 \\              
            &$10<E<100$ & 52    &        5 &  0.11 &      0.15 &  0.32 \\
\hline                                                                                   
FSRQ        &$E>0.1$    & 94	&	9 &   0.30 &      0.38 & $4.5\times 10^{-4}$ \\
            &$0.1<E<0.3$& 93	&	9 &   0.34 &      0.41 & $1.9\times 10^{-4}$ \\
\enddata
\end{deluxetable}

\clearpage

\clearpage

\end{document}